%% file: paper_A07_arXivRevised.tex
\documentclass[11pt,epsbox]{article}
\pdfoutput=1
\usepackage{graphicx}
\usepackage{amsmath,amssymb}
\baselineskip=\normalbaselineskip
\renewcommand{\baselinestretch}{1.4}
\newcommand{\resection}[1]
 {\setcounter{equation}{0}\section{\large{#1}}}

\begin{document}

\input{promise}

\setcounter{page}{0}
\begin{flushright}
March 2016 
\end{flushright}

\vfill

\begin{center}
{\Large{\bf 
Regular Black Holes \\ and \\
Noncommutative Geometry Inspired Fuzzy Sources
}
}
\end{center}

\vfill

\renewcommand{\baselinestretch}{1.0}

\begin{center}
{\sc Shinpei Kobayashi}
\footnote{e-mail: {\tt shimpei@u-gakugei.ac.jp}}  

~\\

$^{1}${\sl Department of Physics, Tokyo Gakugei University, \\ 
     4-1-1 Nukuikitamachi, Koganei, Tokyo 184-8501, JAPAN } \\

\end{center}

\vfill

\begin{center}
{\bf abstract}
\end{center}

\begin{quote}

\small{%
We investigated regular black holes with fuzzy sources 
in three and four dimensions. 
The density distributions of such fuzzy sources 
are inspired by noncommutative geometry 
and given by Gaussian or generalized Gaussian functions. 
We utilized mass functions to give a physical interpretation 
of the horizon formation condition for the black holes. 
In particular, we investigated three-dimensional BTZ-like black holes
and four-dimensional Schwarzschild-like black holes in detail, 
and found that the number of horizons
is related to the spacetime dimensions, 
and the existence of a void in the vicinity 
of the center of the spacetime is significant,  
rather than noncommutativity. 
As an application, we considered a three-dimensional 
black hole with the fuzzy disc  
which is a disc-shaped region known in the context of 
noncommutative geometry as a source.
We also analyzed a four-dimensional black hole
with a source whose density distribution is an extension 
of the fuzzy disc, and investigated the horizon formation condition 
for it. 
}
\end{quote}
\vfill

\renewcommand{\baselinestretch}{1.4}

\renewcommand{\thefootnote}{\arabic{footnote}}
\setcounter{footnote}{0}
\addtocounter{page}{1}
\newpage

\resection{Introduction}

Quantum features of spacetime have been discussed for a long time
and there have been so many trials to depict their physics. 
Instead of enthusiastic studies, we still do not know 
which manner can be the most natural criterion 
to quantize a spacetime. 
Consequently, our starting point is to investigate such 
a phenomenon that definitely appears when a spacetime is 
consistently quantized. 
Spacetime noncommutativity is one of such features,   
even though there might be diverse ways to impose 
noncommutativity\cite{Aschieri:2005zs, Aschieri:2005yw, 
Aschieri:2009qh, Asakawa:2009yb, Kobayashi:2009baa}. 
We can naively expect that noncommutativity makes 
various changes in the structures of spacetimes, 
in particular, black hole spacetimes 
and the very early universe. 

In this context, the authors of \cite{Nicolini:2005vd} investigated a four-dimensional spacetime
with a source inspired by noncommutative geometry. 
As a consequence of noncommutativity,  
they proposed a source that has a Gaussian distribution $e^{-r^2/(2\theta)}$ 
instead of a delta function $\delta^{(3)}(\vecvar{r})$
since we have to abandon a picture of zero size object like a point particle 
and should replace it by something smeared.  
Here $\theta$ is a noncommutative parameter that represents
spacetime noncommutativity, e.g., $[x, y]=i\theta$ in a two-dimensional space. 
They found that there can exist a black hole with such a source at its center. 
It is a regular black hole in the sense that the curvature singularity at the center 
is resolved since the matter source is diffused by noncommutativity. 
What we want to focus on here about the black hole in \cite{Nicolini:2005vd} 
is that it can have two horizons 
as long as an appropriate condition 
is satisfied, even though it is not charged, nor does it have an angular momentum. 
The existence of a black hole with two horizons 
means that there would be an extreme black hole 
where two horizons coincide 
and there would appear a remnant after the Hawking radiation 
starting from a non-extreme black hole. 
This may change a story of the black hole evaporation. 
Inspired by this fascinating scenario, a lot of works on black holes 
with such Gaussian sources have been done so far 
\cite{Ansoldi:2006vg, Nicolini:2009gw, Nicolini:2011fy, Spallucci:2008ez, 
Smailagic:2010nv, Modesto:2010rv, Nicolini:2008aj, Mureika:2011py, Larranaga:2014uca}. 

One more thing we want to note is that 
there is always a solution 
for any density distribution because the corresponding 
energy-momentum tensor of anisotropic fluid 
compensates for a consistent solution to exist. 
This is another reason that many authors have been able to consider 
these noncommutative geometry inspired black holes, 
which also has been referred in the context of another type of regular black holes 
\cite{Dymnikova:1992ux, Dymnikova:2003vt, Dymnikova:2004qg}. 

It is thereby natural that this  research has been extended 
to three-dimensional black holes. 
Though a three-dimensional spacetime is intrinsically 
different from a four-dimensional spacetime, 
as is well known, there exists the BTZ black hole in a three-dimensional 
spacetime with a negative cosmological constant. 
The authors of \cite{Rahaman:2013gw, Tejeiro:2010gu, Larranaga:2010tt, 
Rahaman:2014pha, Myung:2008kp} 
analyzed three-dimensional black holes with Gaussian sources
which have similar structures to the BTZ black hole, 
in the sense that the spacetimes are asymptotically anti de Sitter, 
but at the same time, 
there are de Sitter cores around the centers on the contrary to the BTZ black hole.  
As we will see later, 
there is no black hole with two horizons due to the core \cite{Myung:2008kp}. 
Motivated by this fact, the authors of  \cite{Myung:2008kp} and 
\cite{Jun:2014jqa} introduced generalized Gaussian sources
whose density distributions are proportional to $r e^{-r^2/(2\theta)}$
and $r^2 e^{-r^2/(2\theta)}$, respectively.
The change of sources makes a black hole 
have two horizons as long as an appropriate condition is satisfied. 

The aim of this paper is to clarify what is physically essential for a spacetime  
with such a fuzzy source to have a horizon.
In particular, we are interested in how noncommutativity changes 
the number of horizons. 
Here we want to move away from the specifics and consider general properties. 
To this end, we utilize a mass function that denotes the mass within a given radius. 
Such a mass function determines the condition for a spacetime to have a horizon,
since the necessary mass that must be included within the horizon radius 
is automatically determined, once the radius of a black hole is given. 

In the rest of this paper,  we will investigate 
the existence of horizons and the number of them 
for a three-dimensional black hole with a source 
described by a generalized Gaussian $r^n e^{-r^2/(2\theta)}$, 
using a mass function and a characteristic function 
which denotes the horizon formation condition. 
In order to do so, we will solve the Einstein equation with 
anisotropic fluid corresponding to the source and the 
negative cosmological constant. 
Also, we will see that, for a three-dimensional black hole, 
the existence of a void around its center is crucial to have two horizons. 
We use a toy model whose density distribution 
is not related to noncommutativity to check our statement. 

Since the characteristic function we will propose here 
to judge the horizon formation is intuitive and graphically versatile, 
we can apply it to various cases. 
In fact, we consider a three-dimensional black hole
with a source whose density distribution is originally motivated by the fuzzy disc
in noncommutative geometry. 
The fuzzy disc is a disc-shaped region in a two-dimensional Moyal plane
and its corresponding function is a sum of density distributions 
represented by the generalized Gaussian functions. 
We will also investigate an extension of the density distribution
of the fuzzy disc type and a black hole around it in a four-dimensional spacetime. 

This paper is organized as follows. In Sec.2, we show how a mass function
is used to determine the horizon formation condition, 
using the Reissner-Nortstr{\o}m black hole as an example,  
and we apply the same manner 
to the four-dimensional black hole argued in \cite{Nicolini:2005vd}.  
In Sec.3,  we will analyze three-dimensional black holes with fuzzy sources 
whose density distributions are given by the generalized Gaussian functions. 
We will investigate the characteristic function 
for the horizon formation condition in detail,
and will see what is essential for a horizon to be formed. 
In Sec.4, noncommutative geometry inspired black holes with 
sources motivated from the fuzzy disc are considered.  
Sec.5 is devoted to conclusion and discussion. 
We also refer a black hole spacetime with multi-horizon
and the fuzzy annulus as its source.

\resection{Mass function and horizon formation condition}

The existence of a black hole, in other words, the existence of a horizon,  
depends on how much mass is condensed in a given region. 
Even if there is a large amount of mass, but it is too diffused, 
a black hole horizon can not be formed. 
Since the sources we will treat in this paper are smeared by 
replacing the delta function to the Gaussian functions, 
how much mass exists within a given radius is essential 
for a spacetime to have a horizon. 
A mass function is intuitively useful to express such a necessary mass. 

\subsection{Reissner-Nortstr{\o}m black hole and 
horizon formation condition}

In order to judge when a horizon is formed for a noncommutative 
geometry inspired black hole, we can utilize a mass function.  
It is the profile of the mass distribution that is calculated by the volume 
integration of a density. 
It also can be regarded as an effective mass that is obtained 
by an analogue to 
the Schwarzschild mass. 
For example, let us consider the Reissner-Nortstr{\o}m (RN) solution.
In $G=c=1$ unit, the line element of the four-dimensional 
RN black hole is given by
\be
ds^2 = -\left(1-\frac{2M}{r}+\frac{Q^2}{r^2}\right)dt^2
+\left(1-\frac{2M}{r}+\frac{Q^2}{r^2}\right)^{-1}dr^2
+r^2 d\Omega_{(2)}^2,
\ee
where $M$ is the total mass in the spacetime, 
and $Q$ is the electric charge of the black hole. 
The existence of a horizon 
is determined by the divergent behavior of the $(rr)$-component 
of the metric. In other words, the number of roots for the equation 
\be
f(r) = 1-\frac{2M}{r}+\frac{Q^2}{r^2} = 0,
\label{RNhorizon}
\ee
corresponds to the number of the horizons. 
If $M > |Q|$, the RN metric describes the black hole spacetime 
with two horizons. They are located at $r_{\pm} = M \pm \sqrt{M^2-Q^2}$. 
If $M=|Q|$, there is a special type of a black hole with one horizon. 
This is an extreme RN black hole in which $r_+$ and $r_-$ coincide. 
If $M < |Q|$, there is no black hole, but a naked singularity 
that does not have a horizon.

We can graphically clarify if a horizon exists or not 
by introducing a mass function. 
The mass function $m(r)$  
for the RN black hole is defined as
\be
m(r) = M -\frac{Q^2}{2r}.
\ee
Using $m(r)$, the line element of the RN black hole 
is rewritten as 
\be
ds^2 = -\left(1-\frac{2m(r)}{r}\right)dt^2
+\left(1-\frac{2m(r)}{r}\right)^{-1}dr^2
+r^2 d\Omega_{(2)}^2,
\ee
which can be regarded as 
the line element of the Schwarzschild black hole with the effective mass $m(r)$. 
Eq.(\ref{RNhorizon}) is also rewritten as 
\be
f(r) = 1-\frac{2m(r)}{r} = 0,
\ee
which gives the horizon formation condition. 

One of the advantages of this perspective is 
that it enables us to understand 
why an infalling observer can avoid hitting the singularity 
at the center of the RN black hole. 
Actually, inside the inner horizon 
($0 \leq r \leq r_-$), the mass function is always negative, 
which makes the gravitational force effectively repulsive there \cite{Poisson}. 

Another advantage of introducing the mass function, which will become
more significant in the following analyses in this paper,  
is that it makes us possible to argue
the horizon formation condition based on the analogue of a well-known 
black hole. 
Though there is no geometrical basis to define a mass function, 
we can choose a simpler and more useful one 
for a spacetime we want to consider. 
Clearly, for four-dimensional regular black holes, 
we can use the Schwarzschild black hole as such. 
The Schwarzschild horizon depends on its mass as
\be
r_h = 2M,
\ee
which means that if there is a Schwarzschild black hole with radius $r_h$, 
the total mass $M_h = r_h/2$ must be included 
within radius $r_h$. 
More precisely, if a mass included inside a sphere of radius $r_h$ 
is equal to or larger than $r_h/2$, 
a black hole is formed. 
 
\begin{figure}[tb]
  \begin{center}
    \includegraphics[scale=0.6, bb=0 0 300 214]{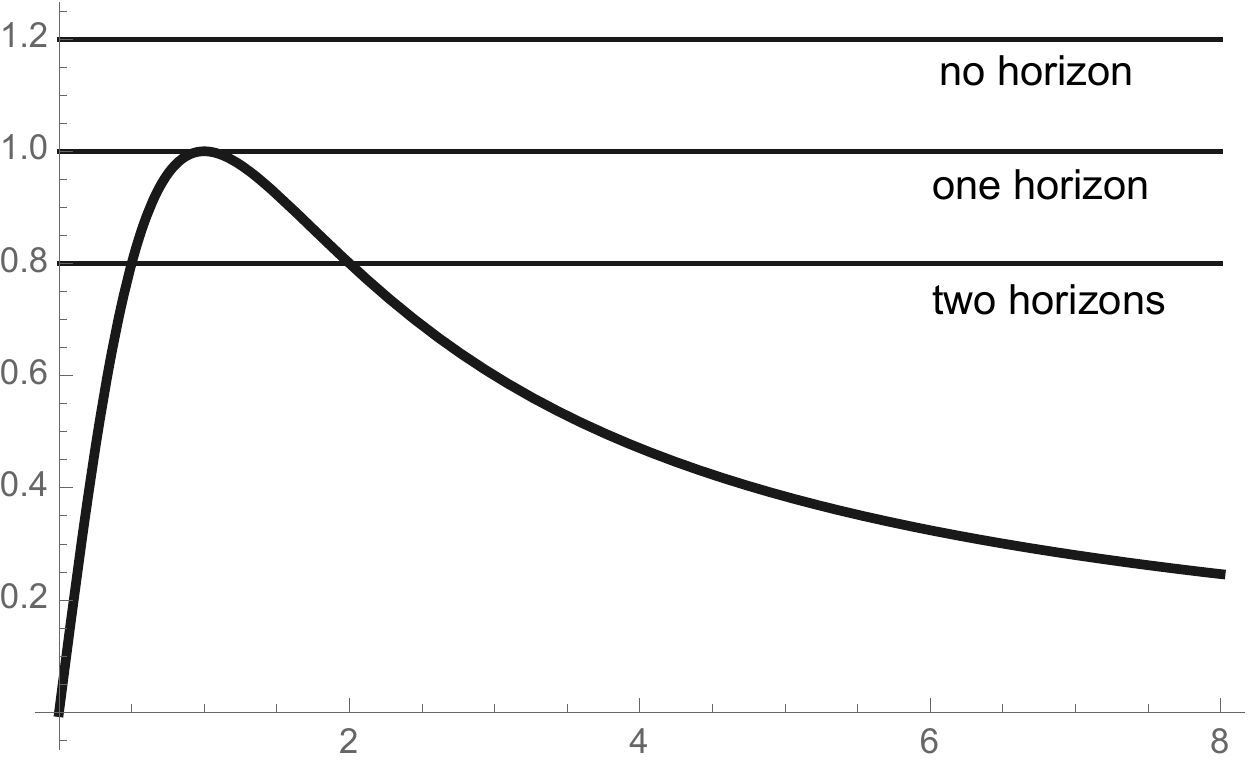}
    \caption{Plot of $h(x)$ with $q=1$ defined in (\ref{RNhorizonfunc}). 
    The horizontal lines denote the values of $L/M$. 
    The number of intersections corresponds to the numbers of 
    horizons.} 
    \label{RNhorizonFig}
  \end{center}
\end{figure}

Applying this idea to the RN case, we can interpret the horizon formation condition 
using the mass function as the existence of $r_h$ that satisfies
\be
m(r_h) \geq M_h = \frac{r_h}{2}.
\ee
This condition states that once a horizon radius is given, 
the total mass that must be included within the radius 
will be determined automatically. 
Of course, this condition obviously coincide with $f(r)=0$, 
but our point of view is physically more apparent. 
For the RN case, the condition for a horizon to be formed can be rewritten as
\be
M-\frac{Q^2}{2r_h} \geq \frac{r_h}{2} \quad \Leftrightarrow \quad
\frac{1}{M} \leq \frac{2r_h}{r_h^2 + Q^2}.
\label{HorizonCondiRN}
\ee
The existence of a horizon is determined by the number of 
intersections between the following characteristic function 
\be
h(x)=\diss \frac{2x}{x^2 + q^2}, 
\label{RNhorizonfunc}
\ee
and the constant function that represents the value of $L/M$. 
Here $L$ is a typical length of this spacetime, 
which is introduced to define
dimensionless parameters $L/M$, $x \equiv r_h/L$ and $q \equiv Q/L$. 
Now the condition (\ref{HorizonCondiRN}) is translated to 
\be
\frac{L}{M} \leq h(x)=\diss \frac{2x}{x^2 + q^2}. 
\ee
The profile of $h(x)$ is shown in Fig.\ref{RNhorizonFig}. 
By the way, we are writing the condition for the inverse of mass $1/M$, 
not for $M$. 
This is because $h(x)$ does not diverge both around $x=0$ and 
for $x \to \infty$, which makes it simpler 
to analyze the behavior of the horizon formation condition 
around $x=0$ and $x\to \infty$.  
This usage of the mass function have not been seen in the previous works. 

The characteristic function $h(x)$ takes its maximum value $1/q$ at $x= q$. 
$L/M$ must therefore be equal to or smaller than $1/q$ 
in order that at least one horizon exists. 
When $M=Q\  (\Leftrightarrow L/M=1/q)$, 
there is one horizon, which corresponds to the extreme black hole.
For $M > Q$, there are two horizons. 
These facts on the RN black hole are well known. 

We will investigate the horizon formation conditions for various 
sources in the same manner in the rest of this paper. 
As mentioned before, there is no generally natural definition 
of a mass function for an arbitrary spacetime, 
and we can use an suitable form for a spacetime we want to consider. 
In fact, we will consider an analogue of the BTZ black hole 
to define a mass function in three dimensions, 
on the contrary to the Schwarzschild black hole in 
four dimensions.\footnote{
Furthermore, we can choose different types of mass 
functions if a black hole is charged and/or rotating, 
similar to the RN black hole. 
}

\subsection{Horizon formation condition for a four-dimensional 
noncommutative geometry inspired Schwarzschild black hole}

We want to apply the method in the previous subsection
to investigate the horizon formation condition 
for a four-dimensional 
regular black hole inspired by noncommutative geometry
considered in \cite{Nicolini:2005vd}. 
The density distribution of the source of the black hole has a Gaussian shape\footnote{
$\theta$ in this paper is twice as large as the one used in \cite{Nicolini:2005vd}. 
}
 $\rho(r) \propto e^{-r^2/(2\theta)}$. 
Here $\theta$ is a noncommutative parameter 
that defines the canonical commutation relation 
between space coordinates as 
\be
[x, y] = i\theta.
\ee
When this relation is imposed to a space, we can naively expect that 
there is no `zero-size' object.
For example, a source of the delta function type would be smeared and fuzzy.
Then one of the simplest realizations is to replace 
the delta function to a Gaussian function
\be
\delta^{(3)}(\vecvar{r}) \to \exp\left(-\frac{r^2}{2\theta}\right). 
\ee 
The authors of \cite{Nicolini:2005vd} made use of the fact 
that for any density distribution, there exist the corresponding 
solution for the Einstein equation because of compensating 
by an appropriate component of the energy-momentum tensor
of anisotropic fluid. 
In \cite{Nicolini:2005vd}, the tangential pressure $T_{\phi\phi}$
plays the role. 
This has been extend to various black holes, e.g., charged \cite{Ansoldi:2006vg}, 
rotating \cite{Smailagic:2010nv, Modesto:2010rv}, 
or lower-dimensional \cite{Mureika:2011py} and higher-dimensional ones \cite{Spallucci:2009zz}, 
and so on. The reference \cite{Nicolini:2008aj} is a review of noncommutative geometry 
inspired black holes written by one of the authors of \cite{Nicolini:2005vd}. 

The solution shown in \cite{Nicolini:2005vd} is given by 
\be
ds^2 = -\left(1-\frac{2m_{4d}(r)}{r}\right)dt^2 
+\left(1-\frac{2m_{4d}(r)}{r}\right)^{-1}dr^2 
+r^2\left(d\theta^2 + \sin^2\theta d\phi^2\right),
\ee
where 
\ba
m_{4d}(r) &=& 4\pi \int_0^r dr' r'^2 \rho_{4d}(r') 
= 4\pi\int_0^r dr' r'^2\frac{M}{(2\pi\theta)^{3/2}} \exp\left(-\frac{r'^2}{2\theta}\right)  
\nonumber\\
&=& \frac{2M}{\sqrt{\pi}}
\gamma\left(\frac{3}{2}, \frac{r^2}{2\theta}\right),
\ea
is the mass function for this system.\footnote{
For a more general profile of density \cite{Nicolini:2011fy}
\be
\rho_{4d}(r) = \frac{M}{4\pi\theta(2\theta)^{\frac{n+1}{2}}
\Gamma\left(\frac{n+3}{2}\right)}
r^n \exp\left(-\frac{r^2}{2\theta}\right), 
\ee
the corresponding mass function  is given by
\be
m_{4d}(r) = 4\pi \int_0^r dr' r'^2 \rho(r') 
= \frac{M}{\Gamma\left(\frac{n+3}{2}\right)}
\gamma\left(\frac{n+3}{2}, \frac{r^2}{2\theta}\right).
\ee
We can apply the manner in this paper 
to this generalized distribution. 
}
$\gamma(a,x)$ is the lower incomplete gamma function 
related to the upper incomplete gamma function as 
\be
\gamma(a, x) = \Gamma(a) -\Gamma(a, x). 
\ee
The normalization is determined by $m_{4d}(r=\infty)=M$, 
which gives the total mass in the whole space. 

Repeating the same argument for the RN black hole, 
we can interpret the horizon formation condition as the existence of $r_h$ that satisfies
\be
m_{4d}(r_h) \geq M_h = \frac{r_h}{2} 
\quad \Leftrightarrow \quad 
\frac{2M}{\sqrt{\pi}}
\gamma\left(\frac{3}{2}, \frac{r^2}{2\theta}\right) 
\geq \frac{r_h}{2}. 
\ee
\begin{figure}[tb]
  \begin{center}
    \includegraphics[scale=0.6, bb=0 0 350 214]{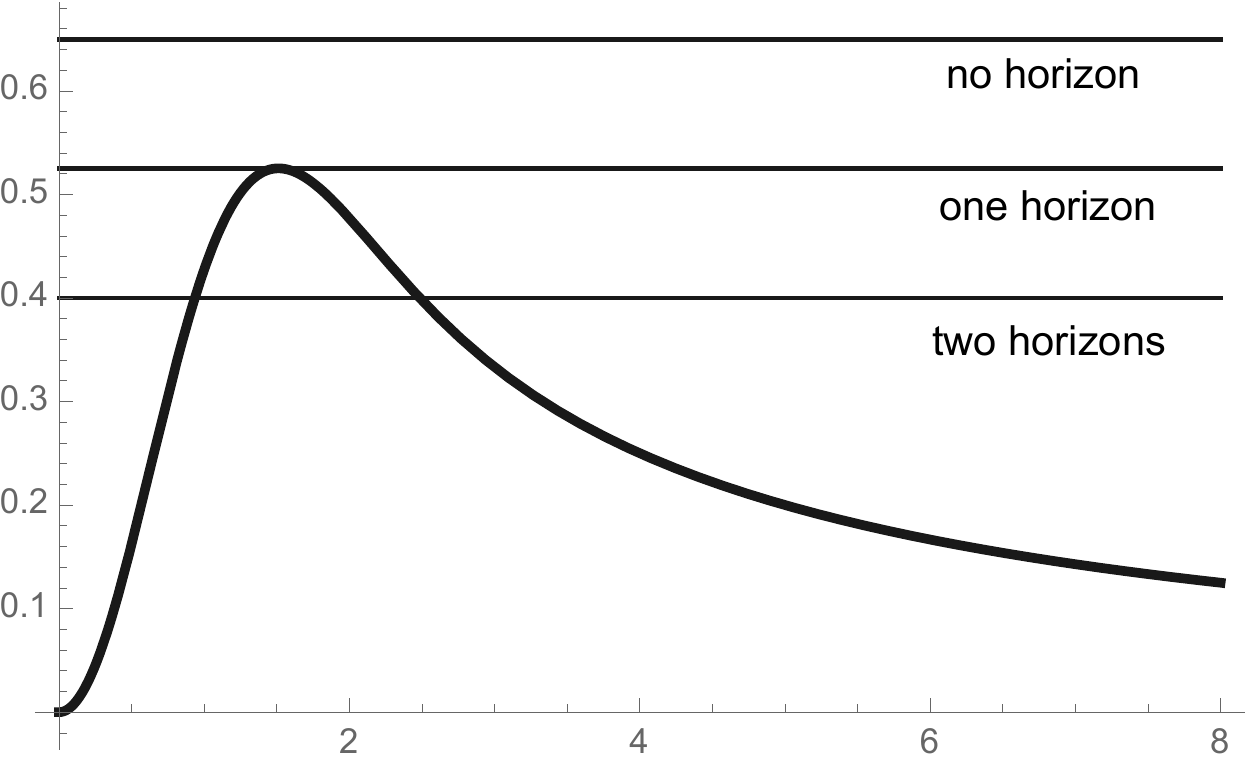}
    \caption{Plot of $h_{4d}(x)$. 
    The horizontal lines denote the values of $\sqrt{2\theta}/M$. 
    $h_{4d}(x)$ takes the maximum value $h_{4d}^* = 0.525$ at $x=1.51$. 
    When $M > \sqrt{2\theta}/h_{4d}^*$, there are two horizons. 
    When $M = \sqrt{2\theta}/h_{4d}^*$, there is one horizon, which corresponds 
    to the extreme black hole. When $M < \sqrt{2\theta}/h_{4d}^*$, 
    there is no horizon, which means that no black hole is formed
    but a regular lump of mass like a star exists. } 
    \label{NicoliniCondi}
  \end{center}
\end{figure}
Introducing a dimensionless parameter $x = r_h/\sqrt{2\theta}$, 
the condition is interpreted as the existence of $x$ that satisfies
\be
h_{4d}(x)
\equiv
\frac{2\gamma\left(\frac{3}{2}, x^2\right)}{\sqrt{\pi}x}
\geq \frac{\sqrt{2\theta}}{M}.
\label{h4d}
\ee
The plot of $h_{4d}(x)$ is shown in Fig.\ref{NicoliniCondi}.
$h_{4d}(x)$ takes the maximum value $\thickapprox 0.525$ at $x=1.51$,
which means that the extreme black hole exists when
\be
M \thickapprox \frac{\sqrt{2\theta}}{0.525}. 
\ee
For $M$ which is larger than $\sqrt{2\theta}/0.525$, there is a black hole with two horizons. 
This result coincides with \cite{Nicolini:2005vd}, 
multiplying two to $\theta$ to replace the noncommutative 
parameter used in  \cite{Nicolini:2005vd}.

\resection{Three-dimensional black hole with fuzzy source}

\subsection{Three-dimensional rotating regular black hole with anisotropic fluid}

We can easily extend the analysis in the previous section 
to three-dimensional cases. 
To begin with, let us show that there is a black hole which corresponds to 
any density distribution $\rho(r)$ also in three dimensions. 
\begin{table}[tb]
\begin{center}
\begin{tabular}{ccccc}\hline
$\mbox{}$ & $n$ & $Q$ & $J$ & $\Lambda$ \\ \hline
\hline
Rahaman {\it et al.} \cite{Rahaman:2013gw}
 & $ \quad \ \quad  0 \quad \ \quad $
 & $ \quad \ \quad  0 \quad \ \quad $
 & $ \quad \ \quad  0 \quad \ \quad $
 & $ \quad \ \quad  <0 \quad \ \quad $  \\ \hline
Tejeiro \& Larranaga \cite{Tejeiro:2010gu} & 0 & 0 & \mbox{}  & $< 0$ \\ \hline
Tejeiro \& Larranaga \cite{Larranaga:2010tt} 
& 0 & $\propto e^{-r^2/(4\theta)}$  & 0  & $< 0$ \\ \hline
Rahaman {\it et al} \cite{Rahaman:2014pha}
& 0 & $\propto r^{n+2}$  & 0  & $< 0$ \\ \hline
Myung \& Yoon \cite{Myung:2008kp} 
& 0, 1 & 0 & 0 & $<0$ \\ \hline
Liang, Liu \& Zhu \cite{Jun:2014jqa} 
& 2 & 0 & \mbox{} & $< 0$ \\ \hline
Park \cite{Park:2008ud} 
& $n$ & 0 & 0 & $> 0$ \\ \hline
\end{tabular}
\caption{Three-dimensional black holes with fuzzy sources}
\label{BHs} 
\end{center}
\end{table}
As summarized in Table.\ref{BHs}, various types of three-dimensional, 
noncommutative geometry inspired black holes have been proposed so far. 
All of them were modifications of the BTZ black hole
by replacing densities of the delta function type to the Gaussian type
\be
\rho(r) \propto e^{-\frac{r^2}{2\theta}},
\ee
or the generalized Gaussian type
\be
\rho(r) \propto r^n e^{-\frac{r^2}{2\theta}} \quad (n \geq 1).
\ee

To consider a concrete spacetime, 
we first derive a three-dimensional, circular symmetric solution of 
the Einstein equation with the negative cosmological constant
\begin{eqnarray}
&& G_{\mbox{}\ \nu}^{\mu} = 8\pi T_{\mbox{}\ \nu}^{\mu} 
+ \Lambda\delta_{\mbox{}\ \nu}^{\mu}.
\end{eqnarray}
$\Lambda$ is the cosmological constant which is related to 
the curvature length $\ell$ as $\Lambda = -1/\ell^2$.
In this paper we use $c=G_3=1$ unit, where  
$G_3$ is the three-dimensional gravitational constant.

We want to consider a circular symmetric spacetime described by the following metric  
\begin{equation} 
ds^2 = -f(r) dt^2 +f^{-1}(r)dr^2 + r^2 \left[d\phi + N_{\phi}(r)dt \right]^2.
\label{metric}
\end{equation}
The spacetime denoted by this metric has an angular momentum 
that is similar to the BTZ black hole.\footnote{
Though the most general form of the metric 
with circular symmetry is given by \cite{Yamazaki:2001ue}
\be
ds^2 = -e^{2\alpha(r)} f(r) dt^2 +f^{-1}(r)dr^2 + r^2 \left[d\phi + N_{\phi}(r)dt \right]^2,
\ee
we focus on the type of metric (\ref{metric}) in this paper for simplicity.  
}
For the energy-momentum tensor 
$T_{\mbox{}\ \nu}^{\mu}$, we impose the following ansatz
\ba
&& T_{\mbox{}\ t}^t = -\rho (r), \quad T_{\mbox{}\ r}^r = p_r (r), 
\quad T_{\mbox{}\ \phi}^{\phi} =p_{\phi}(r) \nn \\
&& T_{\mbox{}\ t}^{\phi} = \sigma_a (r), \quad T_{\mbox{}\ \phi}^t = \sigma_b (r), 
\quad \mbox{others are zero}.
\ea
When we consider a rotating solution, i.e., $N_{\phi} \neq 0$,
the energy-momentum tensor can be no longer diagonal. 
In fact, we can not set $T_{\mbox{}\ t}^{\phi} = \sigma_a (r) =0$ 
to solve the equations of motion consistently, though $\sigma_b$ can be zero
as we will see explicitly. 
This point is not referred in \cite{Tejeiro:2010gu} and \cite{Jun:2014jqa} 
though the existence of the $(\phi, t)$-component of the 
energy-momentum tensor does not affect their conclusions. 
Since $T_{\phi t}$ and $T_{t \phi}$ must be same, we find that 
$\sigma_a$ and $\sigma_b$ obey
\be
\sigma_a = \left(-\frac{f}{r^2}+N_\phi^2\right)\sigma_b +N_{\phi} (p_{\phi} +\rho),
\label{sym}
\ee
which can be used to check the consistency.  

Now we find that the Einstein equation 
$G_{\mbox{}\ \nu}^{\mu} = 8\pi T_{\mbox{}\ \nu}^{\mu} 
+ \Lambda\delta_{\mbox{}\ \nu}^{\mu}$ reduces to
\begin{eqnarray}
2f'+r^2[rN_{\phi}^{\prime 2} +2N_{\phi}(3N_{\phi}^{\prime}+rN_{\phi}^{\prime\prime})]
&=& 4r\left(-8\pi \rho  +\frac{1}{\ell^2} \right), 
\label{eom1}\\
2f'+r^3N_{\phi}^{\prime 2}
&=& 4r\left(8\pi p_r +\frac{1}{\ell^2}\right),
\label{eom2}\\
-3r^2N_{\phi}^{\prime 2} +2f'' -2N_{\phi}(3N_{\phi}^{\prime}+rN_{\phi}^{\prime\prime})
&=&4\left(8\pi p_{\phi} +\frac{1}{\ell^2} \right), 
\label{eom3}\\
r(3N_{\phi}^{\prime} +rN_{\phi}^{\prime\prime}) 
&=& 16\pi \sigma_b
\label{eom4}  \\
N_{\phi}(f'+2r^3N_{\phi}^{\prime 2} -rf'') +(f+r^2N_{\phi}^{2})(3N_{\phi}^{\prime}+rN_{\phi}^{\prime\prime})
&=& -16\pi r \sigma_a
\label{eom5}
\end{eqnarray}
They are $(t,t), (rr), (\phi\phi), (t, \phi)$ and $(\phi, t)$-components of the 
Einstein equation, respectively. The prime denotes the derivative with respect to $r$. 
Besides them, we have to consider the covariant conservation of the energy-momentum tensor 
$T_{\mbox{}\ \mbox{}\ ;\nu}^{\mu\nu}=0$. For $\mu =r$, it gives a non-trivial equation
\begin{multline}
r\left\{
f'(p_r+\rho) +r^2N_{\phi}^{\prime} (\sigma_a -N_{\phi}^{2}\sigma_b)
+N_{\phi} \left[ \sigma_b f' -r^2(p_r+\rho) N_{\phi}^{\prime} \right]
\right\} \\
+f(2p_r -2p_{\phi} -2N_{\phi}\sigma_b -r\sigma_b N_{\phi}^{\prime} +2rp'_r) =0.
\label{eom6}
\end{multline}

There are six equations (\ref{eom1})-(\ref{eom6}) 
and one condition (\ref{sym}) for the symmetry of the energy-momentum tensor 
to determine five unknown functions $f, N_{\phi}, \sigma_a, \sigma_b, p_r$ and $p_{\phi}$. 
In deriving solutions, we will put an ansatz to reduce the totally seven equations into 
six. The redundant equation among the rest six equations is due to 
the Bianchi identity.

When we impose a simple ansatz $\sigma_b =0$, Eq.(\ref{eom4}) can easily be integrated as
\be
N_{\phi}(r) = -\frac{J}{2r^2},
\label{RotPart}
\ee
which coincides with the BTZ case. 
$J$ corresponds to the angular momentum of a black hole. 
Substituting this into the other equations, we see that all the other 
unknown functions $f, \sigma_a, p_r$ and $p_{\phi}$ are determined as 
the functions of the energy density $\rho$
\ba
f(r) &=& -16\pi  \int_0^r dr' \ r'\rho(r') +\frac{r^2}{\ell^2} +\frac{J^2}{4r^2}, \nn \\
&=& -8m(r) +\frac{r^2}{\ell^2}+\frac{J^2}{4r^2}
\label{Mpart}
\\
\sigma_a(r) &=& \frac{J}{2r}\rho'(r) \\
p_{r}(r) &=& -\rho(r), \\
p_{\phi}(r) &=& -(r\rho(r))',
\label{sln}
\ea
where we set an integration constant in $f(r)$ to zero  
for this solution to coincide with the BTZ black hole for a large $r$. 
$m(r)$ is the mass function for a given density $\rho$, which is defined by
\be
m(r) = 2\pi \int dr' r' \rho(r').
\ee
The Ricci scalar for this solution in terms of $f$ and $N_{\phi}$ is 
\be
R = -f''(r) -\frac{2}{r}f'(r)+\frac{1}{2}r^2N_{\phi}^2(r). 
\label{3dimRicci}
\ee
Substituting (\ref{Mpart}) and (\ref{RotPart}) to (\ref{3dimRicci}), 
we obtain
\be
R = 16\pi (3\rho(r)+r\rho'(r)) -\frac{6}{\ell^2} + \frac{J^2}{8r^2}.
\label{Ricci1}
\ee
The energy-momentum tensor with lower indices is given by 
\be
(T_{\mu\nu}) 
= \left(
\begin{array}{ccc}
T_{tt}  & T_{tr}  & T_{t\phi} \\
T_{rt} & T_{rr}  & T_{r\phi}  \\
T_{\phi t} & T_{\phi r}  & T_{\phi\phi}
\end{array}
\right)
=
\left(
\begin{array}{ccc}
f\rho + \diss\frac{J}{2}(r\rho)'  & 0  & \diss\frac{J}{2}(r\rho)' \\
0 & \diss -\frac{\rho}{f}  & 0  \\
\diss\frac{J}{2}(r\rho)' & 0  & -r^2(r\rho)'
\end{array}
\right),
\ee
which is diagonal only for $J=0$ as mentioned before. 

\subsection{Generalized non-Gaussian sources in three dimensions}

We investigate spacetimes with various sources that appear in the context 
of noncommutative geometry.
As an instructive example, let us first see the spacetime with the generalized Gaussian source, 
$\rho \propto r^n e^{-r^2/(2\theta)}$. \footnote{
As summarized in Table.\ref{BHs}, 
the black holes with $n=0, 1$ and with $n=2$ were investigated in  \cite{Myung:2008kp}
and \cite{Jun:2014jqa}, respectively. 
}
To be more concrete, 
we consider the following density distribution described by the generalized Gaussian function
\be
\rho_n(r) = \frac{M}{2\pi\theta (2\theta)^{\frac{n}{2}}\Gamma\left(\frac{n}{2}+1\right)}
r^n e^{-\frac{r^2}{2\theta}}. 
\label{NGD}
\ee
The corresponding mass function is
\ba
m_{n}(r) 
&=& 2\pi \int_0^r r'\rho(r')dr' = \frac{M}{\Gamma\left(\frac{n}{2}+1\right)}
\gamma\left(\frac{n}{2}+1, \frac{r^2}{2\theta}\right) \nn \\
&=& M \left[1-\frac{\Gamma\left(\frac{n}{2}+1, \frac{r^2}{2\theta}\right)}{\Gamma\left(\frac{n}{2}+1\right)}\right].
\label{massfunc}
\ea
Similar to the four-dimensional case, 
the mass function is normalized as $m_n(\infty) = M$ 
using $\Gamma(\frac{n}{2}+1, \infty)=\Gamma(\frac{n}{2}+1)$.
The ratio of $M$ to the noncommutative parameter $\sqrt{2\theta}$
determines the horizon formation condition.  

\subsection{Black holes with a generalized Gaussian source and 
physical interpretation of their horizons}
\label{PhysInterpretation}

Hereafter we set $J=0$ for simplicity, 
but the essence of our analysis does not depend on it 
and we can extend this to the case with nonzero $J$.  
Putting the density distribution (\ref{NGD}) to (\ref{sln})
and setting $J=0$, we obtain
\be
ds^2 = -f_n(r)dt^2 + f_n^{-1}(r)dr^2 +r^2 d\phi^2, 
\label{3dimSln1}
\ee
where
\ba
f_n(r) &=& -8m_n(r) +\frac{r^2}{\ell^2} \nn \\
&=& \frac{M}{\Gamma(n+1)}
\left[
\gamma\left(n+1, \frac{r^2}{2\theta}\right) 
+ \frac{r^2}{2\theta}\gamma\left(n, \frac{r^2}{2\theta}\right) 
\right] +\frac{r^2}{\ell^2}, 
\label{3dimSln2}\\
p_{nr}(r) &=& -\rho_{n}(r) \nn \\
&=& -\frac{M}{2\pi\theta (2\theta)^{\frac{n}{2}}\Gamma\left(\frac{n}{2}+1\right)}
r^n e^{-\frac{r^2}{2\theta}}, 
\label{3dimSln3}\\
p_{n\phi}(r) &=& -(r\rho_n(r))' \nn \\
&=& -\frac{M}{2\pi\theta (2\theta)^{\frac{n}{2}}\Gamma\left(\frac{n}{2}+1\right)}
\left(n+1-\frac{r^2}{\theta}\right)r^n e^{-\frac{r^2}{2\theta}}.
\label{3dimSln4}
\ea

In order to obtain the physical interpretation of  
the three-dimensional spacetime described above, 
let us go back to see the BTZ black hole spacetime. 
The non-rotating BTZ solution 
is represented by the following line element  
\be
ds^2 = -\left(-8M + \frac{r^2}{\ell^2}\right)dt^2
+\left(-8M + \frac{r^2}{\ell^2}\right)^{-1}dr^2
+r^2 d\phi^2.
\ee
The horizon radius is given by 
\be
r_h = \sqrt{8M \ell^2 }, 
\ee
which is determined by $g_{tt}=g_{rr}^{-1}=0$.

As shown in the previous sections, 
we can see this equation as the condition for the mass 
that is necessary for a horizon with radius $r_h$ to be formed. 
In this case 
\be
M = \frac{r_h^2}{8\ell^2},
\ee
is the necessary mass inside a circle with radius $r_h$ 
for the BTZ black hole to have the horizon. 
We use this condition to judge whether the three-dimensional 
black hole with the generalized Gaussian source 
can have a horizon or not. 

For the spacetime described by (\ref{3dimSln1})-(\ref{3dimSln4}), 
the mass function is calculated as (\ref{massfunc}). 
The horizon formation condition is thereby interpreted as the existence of 
$r_h$ that satisfies
\be
m_n(r_h) 
= M \left[1-\frac{\Gamma\left(\frac{n}{2}+1, \frac{r_h^2}{2\theta}\right)}{\Gamma\left(\frac{n}{2}+1\right)}\right]
\geq  \frac{r_h^2}{8\ell^2},
\ee
or equivalently, the existence of $x$ that satisfies
\ba
h_n(x) &\equiv& 
\frac{1}{x^2} \left[1-\frac{\Gamma\left(\frac{n}{2}+1, x^2\right)}{\Gamma\left(\frac{n}{2}+1\right)}\right] 
=\frac{1}{x^2}\frac{\gamma\left(\frac{n}{2}+1, x^2\right)}{\Gamma\left(\frac{n}{2}+1\right)} \nn \\
&\geq& \frac{(\sqrt{2\theta})^2}{8M\ell^2},
\label{Condi3dim}
\ea
where $x=r_h/\sqrt{2\theta}$ as before. 
The maximum value of $h_n(x)$ determines the existence of a horizon. 

The behavior of the characteristic function $h_n(x)$ is very simple because it is just the multiplication of 
$x^{-2}$ and the incomplete Gamma function. 
$h_n(x)$ asymptotically approaches to zero when $x \to \infty$ 
since the upper incomplete gamma function $\Gamma(a, x^2)$ approaches $\Gamma(a)$ 
for $x\to \infty$. 
However, note that there is a difference between $n=0$ and $n\geq 1$ 
in the behaviors of $h_n(x)$ around $x=0$. 
Using the expansion of  the upper incomplete gamma function
\be
\Gamma(a, t) = \Gamma(a) 
+ t^a \left(-\frac{1}{a}+\frac{t}{a+1}-\frac{t^2}{2(a+2)}+\cdots \right), 
\ee
we find that $h_n(x)$ behaves around $x=0$ as 
\be
h_n(x) = \frac{x^n}{\Gamma\left(\frac{n}{2}+2\right)} + O\left(x^{n+2}\right). 
\ee
Therefore we obtain 
\ba
h_n(0) = \begin{cases}
0 & (n =0), \\
1 & (n\geq 1).
\end{cases}
\ea

\begin{figure}[tb]
  \begin{center}
    \includegraphics[scale=0.5, bb=0 0 259 214]{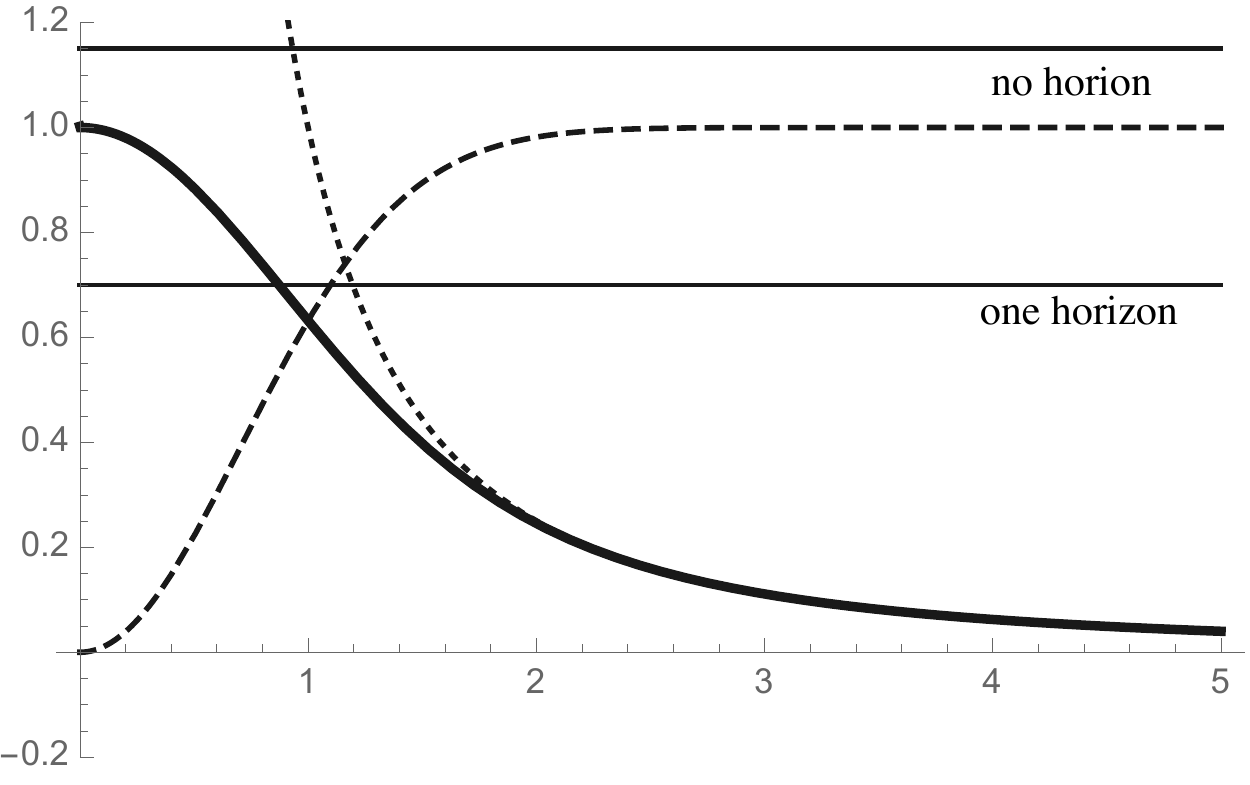}
    \hspace{3cm}
    \includegraphics[scale=0.5, bb=0 0 259 214]{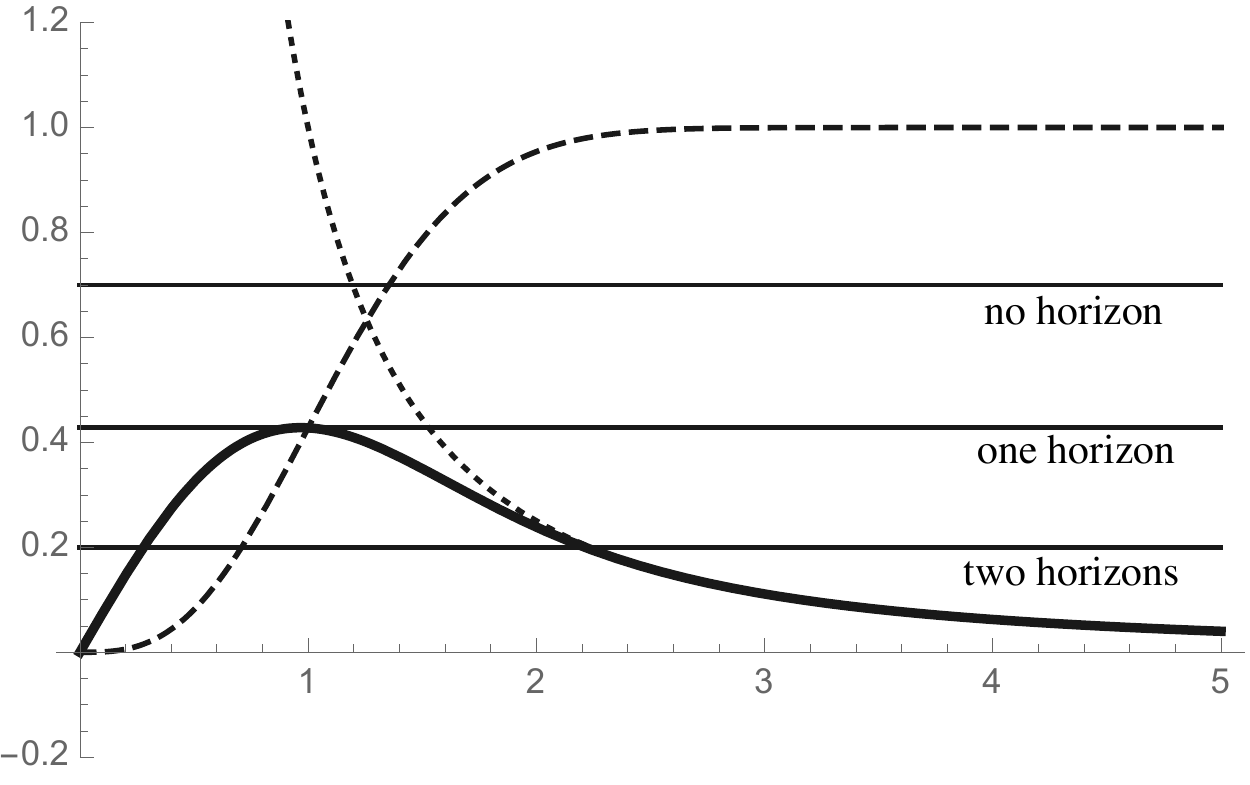}
    \caption{Profiles of $h_0(x)$ (left) and $h_1(x)$ (right). 
    The horizontal lines denote the values of $(\sqrt{2\theta})^2/8M\ell^2$. 
    $h_1(x)$ is chosen as a typical example for $n \geq 1$. 
    For comparison, $x^{-2}$ and $\gamma(n/2+1,x^2)/\Gamma(n/2+1)$ 
    are plotted in the dashed and the dotted curves, respectively.} 
    \label{h_n}
  \end{center}
\end{figure}

Also, $x^{-2}$ is a monotonically decreasing function and diverges at $x=0$, 
on the contrary to the lower incomplete gamma function $\gamma(a, t)$,
which is monotonically increasing and asymptotically approaches to constant. 
Taking the behaviors of both functions 
around $x=0$ into consideration, 
we find that $h_0(x)$ is monotonically 
decreasing with $h_0(0)=1$, 
and $h_n(x)$ for $n \geq 1$ has an extremum at a finite $x$ with $h_n(0)=0$. 
In both cases, $h_n(x)$ asymptotically approaches to zero for $x\to 0$. 
Their behaviors are compared in Fig.\ref{h_n}. 

If the dimensionless constant $(\sqrt{2\theta})^2/8M\ell^2$ is smaller than 
the maximum 
value of $h_n(x)$, there exist a horizon. 
For $n=0$, since $0 \leq h_0(x) < 1$, a horizon is formed when 
\be
0 < \frac{(\sqrt{2\theta})^2}{8M\ell^2} < 1. 
\label{n=0HorCondi}
\ee
If $8M\ell^2/(\sqrt{2\theta})^2 = 1$, there is a ``black hole'' at $r=0$
whose radius is zero.  
So this condition can be read as there must be enough mass 
within a radius for a horizon to be formed  
compared with the noncommutative parameter $\theta$ 
which determines how much the mass is diffused and is leaked out of the radius .  
This is a reasonable claim.\footnote{
In \cite{Rahaman:2013gw} the authors added $M$ to $f_n(r)$
in order to make a spacetime anti-de Sitter around $r=0$
using the ambiguity of integration constant. 
By this modification, the mass function becomes 
\be
m(r) = M\left(\frac{1}{2}-e^{-\frac{r^2}{2\theta}}\right).
\ee
The characteristic function $h_0(x)$ 
diverges negatively for $x \to 0$ and does not take a finite value at $x=0$. 
Therefore $h_0(x)$ in \cite{Rahaman:2013gw} is not a monotonically
decreasing function, but have a maximum at a finite $x$, 
which makes it possible for the black hole to  
have two horizons as long as the mass $M$ is large enough 
compared with the diffusion determined by the noncommutativity 
defined there.
} 

For $n\geq 1$, the horizon formation condition is given by
\be
0 < h_n^{*} \leq 1  
\quad \Leftrightarrow \quad 
M \geq \frac{(\sqrt{2\theta})^2}{8h_n^{*} \ell^2} = M_*,
\ee
where $h_n^*$ is the maximum value of $h_n(x)$. 
When $M > M_*$, there are two horizons. 
The existence of two horizons is one of the peculiar features for $n \geq 1$. 
If $M=M_*$, it is extremal and there exist the black hole with one horizon, 
whose Hawking temperature is zero. 
We can expect that, starting from a state with $M > M_*$, 
the mass $M$ will decrease by the Hawking radiation to the extremal. 
The existence of the extremal state means that there will be a remnant 
after the Hawking radiation even for such a uncharged, non-rotating 
black hole. 
One more difference between $n=0$ and $n\geq 1$ is about the energy conditions 
they satisfy. 
As mentioned in \cite{Nicolini:2005vd} where the four-dimensional case 
with $n=0$ is  considered, the strong energy condition is 
violated for the energy-momentum tensor given in \cite{Nicolini:2005vd}, 
but the weak energy condition is satisfied. 
In the three-dimensional cases, the weak energy condition is 
satisfied in the whole spacetime only for $n=0$. For $n\geq 1$, 
the weak energy condition ($\rho \geq 0$ and $\rho + p_i \geq 0$)
is translated to 
\be
\rho_n \geq 0, \quad \rho_n +p_{nr} \geq 0, \quad \rho_n +p_{n\phi} \geq 0.
\ee
We can explicitly check that 
the first and second condition are always satisfied. 
The third one is rewritten as 
\be
\rho_n +p_{\phi} \propto r^n \left(r^2 -n\theta\right)e^{-\frac{r^2}{2\theta}} \geq 0.
\ee
Therefore $\rho_n+p_{n\phi}$ is not necessarily positive in the whole space. 
We leave the detail analysis and physical meaning of it for a sequent paper. 

Although the existence of the black holes with two horizons for $n \geq 1$
naively appears that the noncommutativity works 
as a repulsive force in the vicinity of the centers of the black holes 
just like the RN case, it is not completely true. 
In fact, in the three-dimensional case we have seen above, 
there is no black hole with two horizons for $n=0$. 
This was pointed out in \cite{Myung:2008kp} and the authors analyzed the 
difference of the behaviors for $n=0$ and $n=1$. 

Actually, the regularity in the whole space 
is realized because of the fuzziness of the source. 
This can be understood by the Ricci scalar at $r=0$. 
Using (\ref{Ricci1}), we can calculate the Ricci scalar at the center of 
the spacetime as
\ba
R|_{r=0}
&=& \begin{cases}
\diss \frac{6}{\ell^2}\left(-1+\frac{8M\ell^2}{(\sqrt{2\theta})^2}\right) & (n=0), 
\vspace{0.3cm}\\
\diss -\frac{6}{\ell^2} & (n \geq 1).
\end{cases}
\ea
For $n\geq 1$, the Ricci scalar becomes a negative constant at $r=0$, 
which is consistent with the fact the mass function with $n \geq 1$ 
is zero at $r=0$ and the negative cosmological constant $\Lambda = -1/\ell^2$ 
is dominant there. 
For $n=0$, there are three cases depending on 
the value of $8M\ell^2/(\sqrt{2\theta})^2$. To be more concrete, 
we find 
\be
R|_{r=0}
= \frac{8M\ell^2}{(\sqrt{2\theta})^2} 
\begin{cases}
> 1 & : \mbox{de Sitter}, \\
= 1 & : \mbox{flat}, \\
< 1 & : \mbox{anti de Sitter}. 
\end{cases}
\ee
As shown in (\ref{n=0HorCondi}), when a horizon is formed, 
$8M\ell^2/(\sqrt{2\theta})^2$ is always larger than 1. 
There is a de Sitter core in the center of the spacetime, 
which is similar to the four-dimensional case \cite{Nicolini:2005vd}.

It is true that in four dimensions, there exist a black hole with two horizons even for $n=0$ 
as long as the mass is large enough. 
To understand the difference between the three- and the four-dimensional 
cases, 
we have to compare the characteristic functions for their horizon formation conditions.  
In the four-dimensional case, the condition is shown in (\ref{h4d}). 
The essential part of the characteristic function is given by
\be
h_{4d}(x) \sim \frac{1}{x}\gamma\left(\frac{n+3}{2}, x^2\right),
\ee
for an arbitrary $n$, where $\sim$ denotes that we are extracting the relevant term. 
On the contrary, in the three-dimensional case, the counterpart is 
given by 
\be
h_n(x) \sim \frac{1}{x^2}\gamma\left(\frac{n+1}{2}, x^2\right). 
\ee
The essential difference is the power of $x$ in front of the lower incomplete 
gamma function that controls the behavior around $x=0$. 
It is clearly originated from the difference of dimensions 
and intrinsic structures due to them 
that appears in $g_{tt} = g_{rr}^{-1}$ 
rather than noncommutativity.

\begin{figure}
 \begin{center}
    \includegraphics[scale=0.5, bb=0 0 259 214]{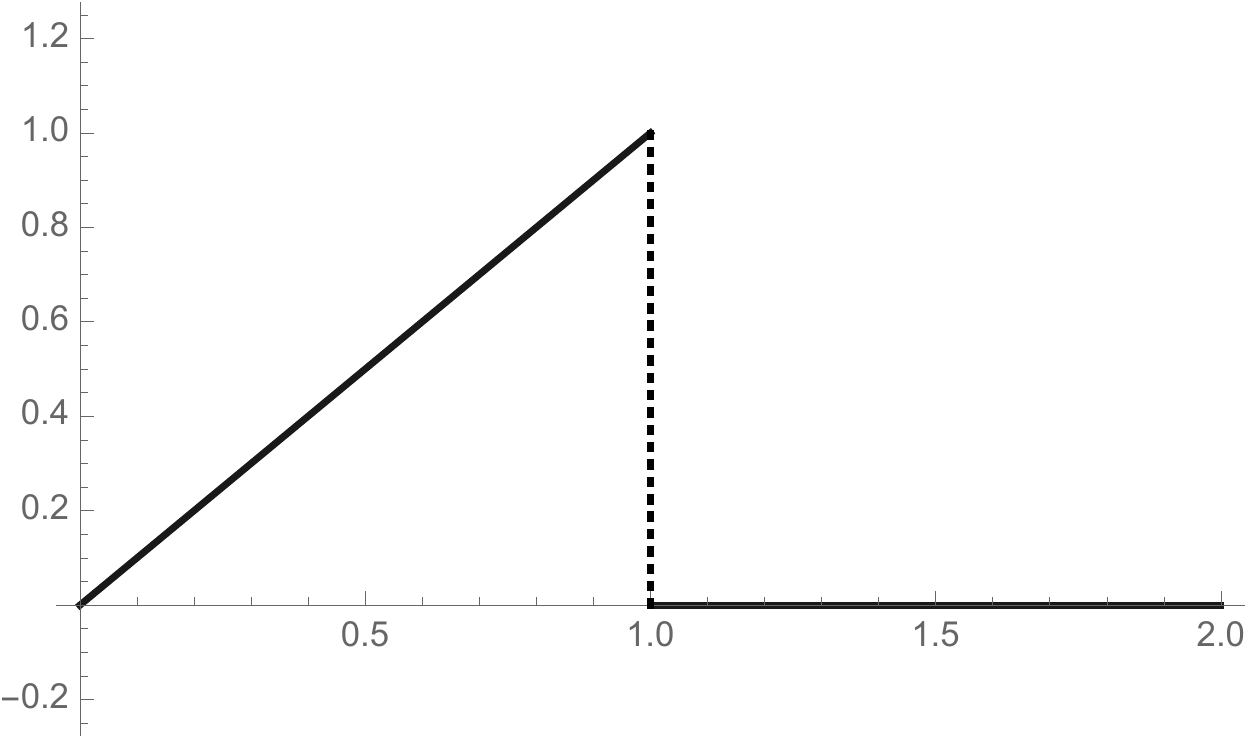}
   \hspace{2cm}
    \includegraphics[scale=0.5, bb=0 0 259 214]{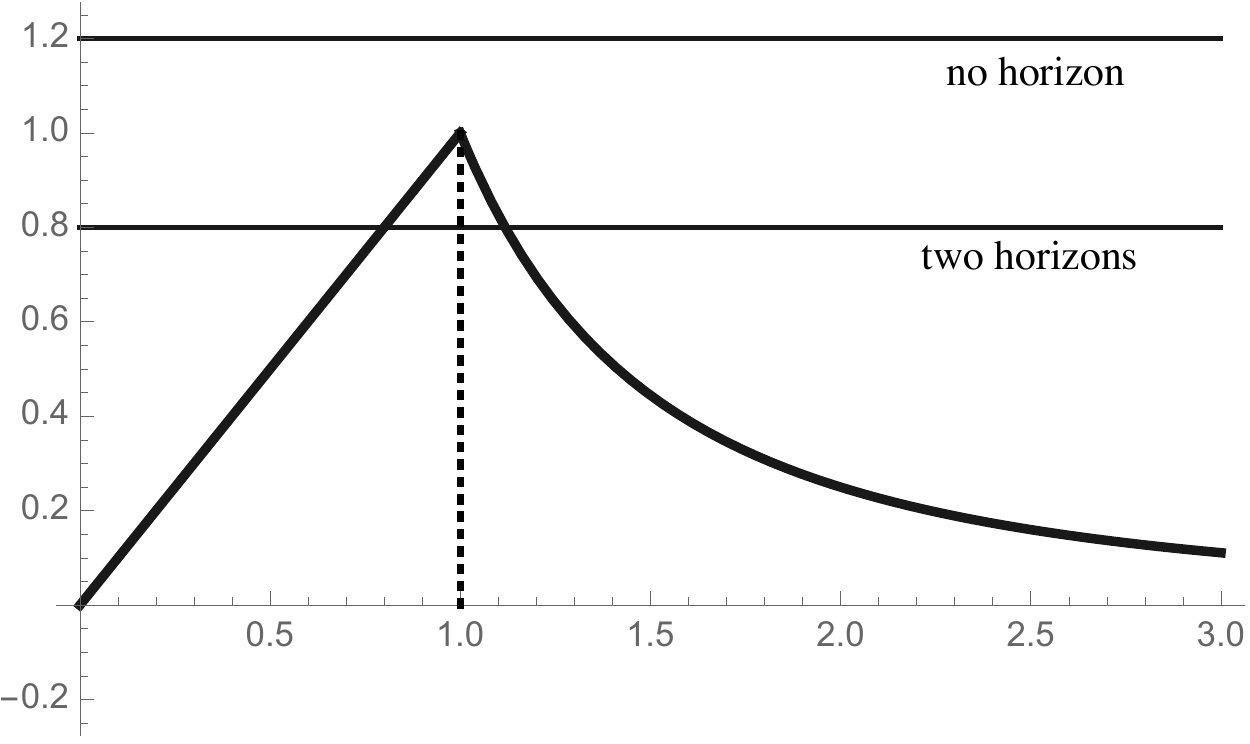}
    \caption{Plots of the toy density (left) and the characteristic function $h_{toy}(y)$ (right).} 
    \label{Toy}
  \end{center}
\end{figure}

To see it more clearly, 
let us consider a simple toy model in three dimensions 
whose density is given by 
\be
\rho(r) =
\begin{cases}
\diss \frac{3M}{2\pi R^3}r & (0 \leq r \leq R), \vspace{0.3cm}
\\
0 & (R < r),
\end{cases}
\ee
where $R$ is a characteristic scale of length of the system. 
The profile of $\rho(r)$ is shown in Fig.\ref{Toy}. 
The mass function for this density is
\be
m(r) = 2\pi\int_0^r dr' r' \rho(r')
= 
\begin{cases}
\diss M\left(\frac{r}{R}\right)^3 & (0 \leq r \leq R), \\
M & (R < r).
\end{cases}
\ee
Note that this model is not realistic in the sense that 
there is a gap the density and the mass function at $r=R$, 
however, it is not crucial in the following argument 
on the existence of a horizon. 
Actually, though we can consider a density that is smooth at $r=R$ 
and has an almost same profile as this toy model, 
it would not give an essential improvement to understand 
the horizon formation condition.  

Then repeating the same argument for the three-dimensional 
black hole with the generalized Gaussian source,
we find that the horizon formation condition is 
given by 
\be
\frac{R^2}{8M\ell^2}
\leq h_{toy}(y) 
\equiv
\begin{cases}
\diss y  & (0 < y \leq 1), \nn 
\vspace{0.3cm}\\
\diss \frac{1}{y^2} & (1 \leq y),
\end{cases}
\ee
where $y$ is a dimensionless parameter defined by $y=r_h/R$. 
The characteristic function for the horizon formation condition 
is shown in Fig.\ref{Toy}. 

The extreme case corresponds to $y=1 \ \Leftrightarrow \ 
R = \sqrt{8M\ell^2}$. For $R < \sqrt{8M\ell^2}$, there are two horizons. 
This existence of two horizons, in particular the existence of 
the inner horizon in this case, 
is a resultant of the void of the mass distribution around the center. 
This implies that there might exist a black hole with two horizons 
as long as there is a void around the center and enough mass is 
condensed in a given region, even if the spacetime noncommutativity 
does not work directly. 

\resection{Regular black hole and fuzzy disc}

\subsection{Fuzzy disc as a source of a three-dimensional black hole}

The analysis so far can be applied to other type of sources inspired from 
noncommutative geometry in three dimensions.
In \cite{Kobayashi:2012ek}, we considered the fuzzy disc, 
which is a disc-shaped region 
in a two-dimensional Moyal plane  \cite{Lizzi:2003ru, 
Lizzi:2003hz, Lizzi:2006bu}.  
A Moyal plane 
is a flat space defined by noncommutative coordinates 
satisfying the commutation relation $[x, y]=i\theta$. 
The algebra of functions on this noncommutative plane is an operator algebra 
$\hat{\cal A}$ generated by
$\hat{x}$ and $\hat{y}$, acting on a Hilbert space
${\cal H}=l^2={\rm span}\{\ket{0},\ket{1},\cdots\}$. 
Here  $\ket{n}$ is an eigenstate of 
``the number operator'' 
\be
\hN \ket{n} = n\ket{n}, \quad \hN \equiv \daga \hat{a},
\ee
defined by the creation and the annihilation operators,  
$\hat{a} = (\hat{x}+i\hat{y})/\sqrt{2\theta},\ 
\hat{a}^{\dagger} = (\hat{x}-i\hat{y})\sqrt{2\theta}$, respectively.

The fuzzy disc is defined by using an operator algebra $\hat{\cal A}$ 
on a Moyal plane
 by restricting to $N\times N$ matrices in the number basis.
It is obtained by the projection 
$\hat{\cal A}_N =\hat{P}_N \hat{\cal A} \hat{P}_N$ 
through the rank $N$ projection operator, 
\be
\hat{P}_{\sss N}= \sum_{n=0}^{N-1}\hat{p}_n =\hat{p}_0 + \cdots + \hat{p}_{\sss N-1},
\label{completenessofP}
\ee
where 
\be
\hat{p}_n= \ket{n}\bra{n} \quad (n=0, 1, \cdots).
\ee

Instead of working with the operators, one can switch to the corresponding 
functions called symbols by means of the Weyl-Wigner correspondence. 
The symbol map based on this correspondence associates 
an operator $\hat{f}$ with a function $f$ as
\be
f(z, \bz) = \bra{z}\hat{f}\ket{z},
\ee
where $z = re^{i\phi}$ and $\ket{z}$ is a coherent state defined by
\be
\hat{a}\ket{z} = \frac{z}{\sqrt{2\theta}}\ket{z}. 
\ee 
Then, 
the corresponding function to the projection operator 
$\hat{p}_n$ is given by
\be
p_n(r) = \bracket{z}{n}\bracket{n}{z}
=e^{-\frac{r^2}{2\theta}}\frac{r^{2n}}{n! (2\theta)^n},
\label{ProjFunc}
\ee
which is one of the realizations of the density distribution described by 
the generalized Gaussian function in the context of noncommutative geometry.
Here we used 
\be
\bracket{z}{n}  = e^{-\frac{r^2}{4\theta}}\frac{\bz^n}{\sqrt{n! (2\theta)^n}}, 
\quad
\bracket{n}{z}  = e^{-\frac{r^2}{4\theta}}\frac{z^n}{\sqrt{n! (2\theta)^n}}.
\ee

One can obtain the corresponding function for the fuzzy disc as well. 
Since the fuzzy disc is a sum of the $N$ projection operators from $n=0$ to $n=N-1$, 
the corresponding function for the fuzzy disc is given by
\be
P_{\sss N} (r) = 
\sum_{n=0}^{N-1} e^{-\frac{r^2}{2\theta}}\frac{r^{2n}}{n! (2\theta)^n}
=\frac{\Gamma (N, \frac{r^2}{2\theta})}{\Gamma (N)}.
\label{incomplete gamma}
\ee  
This function is roughly a radial step function that picks up a disc-shaped region 
around the origin $r=0$ with radius $R=\sqrt{2N\theta}$.
On more details of how to find the corresponding function to an operator,  
one can find in \cite{Kobayashi:2012ek}. 

\begin{figure}[t]
  \begin{center}
   \includegraphics[scale=0.5, bb=0 0 259 214]{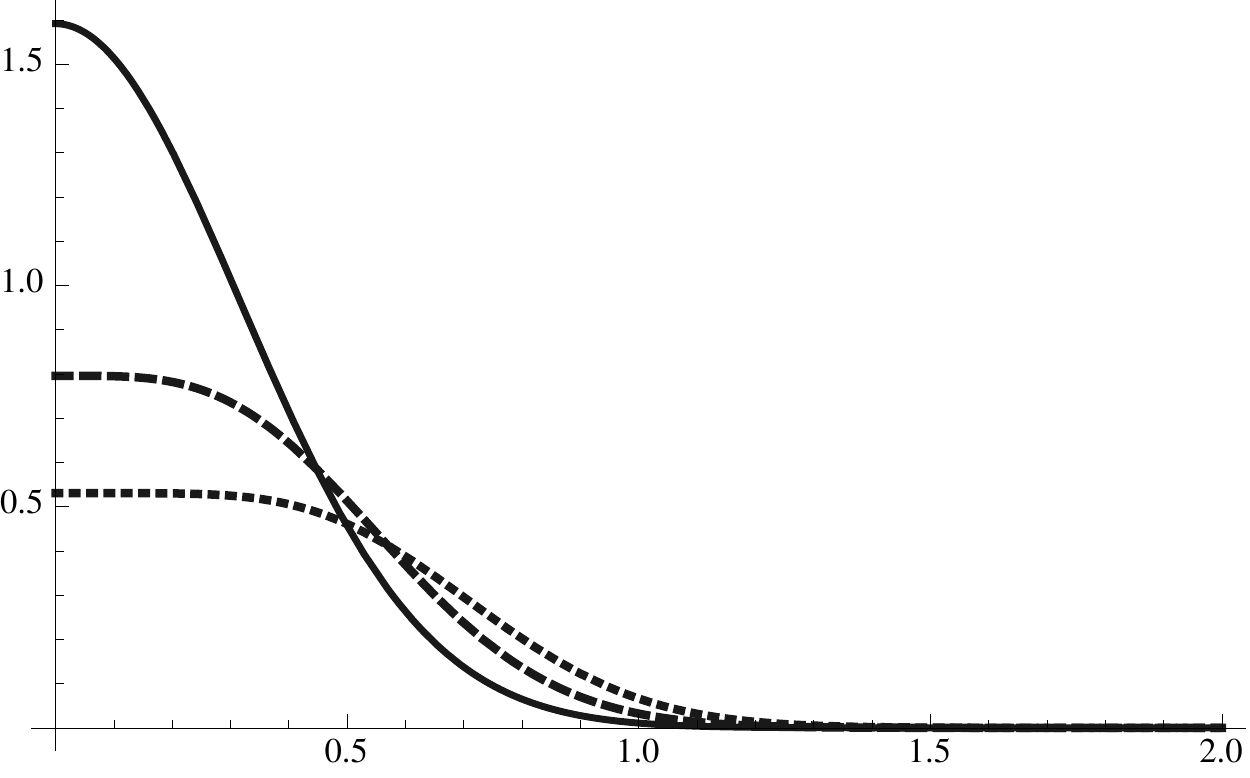}
    \caption{Plot of the density functions of the fuzzy disc type
    for $N=1$ (solid), $N=2$ (dashed) and $N=3$ (dotted). 
    Here we set $M=1$ and $\theta =0.1$. } 
    \label{density}
  \end{center}
\end{figure}

\begin{figure}[t]
  \begin{center}
   \includegraphics[scale=0.5, bb=0 0 300 214]{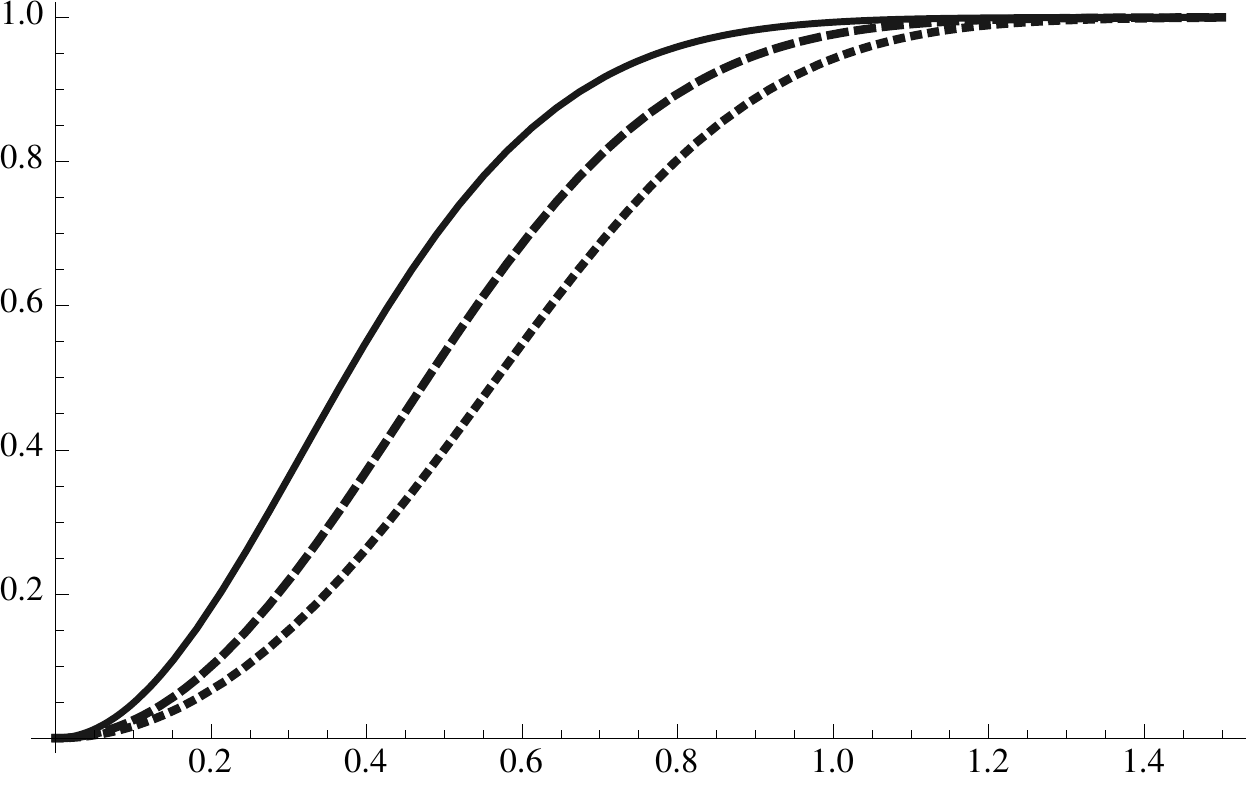}
    \caption{Plot of the mass functions corresponding to the fuzzy disc type 
    source for $N=1$ (solid), $N=2$ (dashed) and $N=3$ (dotted). 
    Here $M=1$ and $\theta =0.1$. } 
    \label{MassFunction}
  \end{center}
\end{figure}

We now use the fuzzy disc as a source for a noncommutative 
inspired black hole in three dimensions. 
As a density motivated by the fuzzy disc (\ref{incomplete gamma}), 
we consider a spacetimes with a density distribution 
\be
\rho_{\sss N}^{\sss FD} (r) = \frac{M}{2\pi \theta N}P_{\sss N}(r)
=\frac{M}{2\pi\theta}\frac{\Gamma (N, \frac{r^2}{2\theta})}{\Gamma (N+1)}. 
\ee
The density distributions (that is, the shapes of the fuzzy discs) 
and 
the mass functions $m_{\sst N}^{\sss FD}(r)$ for $N=1 ,2, 3$ are shown in 
Fig.\ref{density} and Fig.\ref{MassFunction}, respectively. 
They are normalized as 
\be
m_{\sss N}^{\sss FD}(\infty) = 2\pi\int_0^{\infty} dr'  r'\rho_N(r) = M, 
\ee
as before. 

\begin{figure}[t]
  \begin{center}
   \includegraphics[scale=0.4, bb=0 0 259 214]{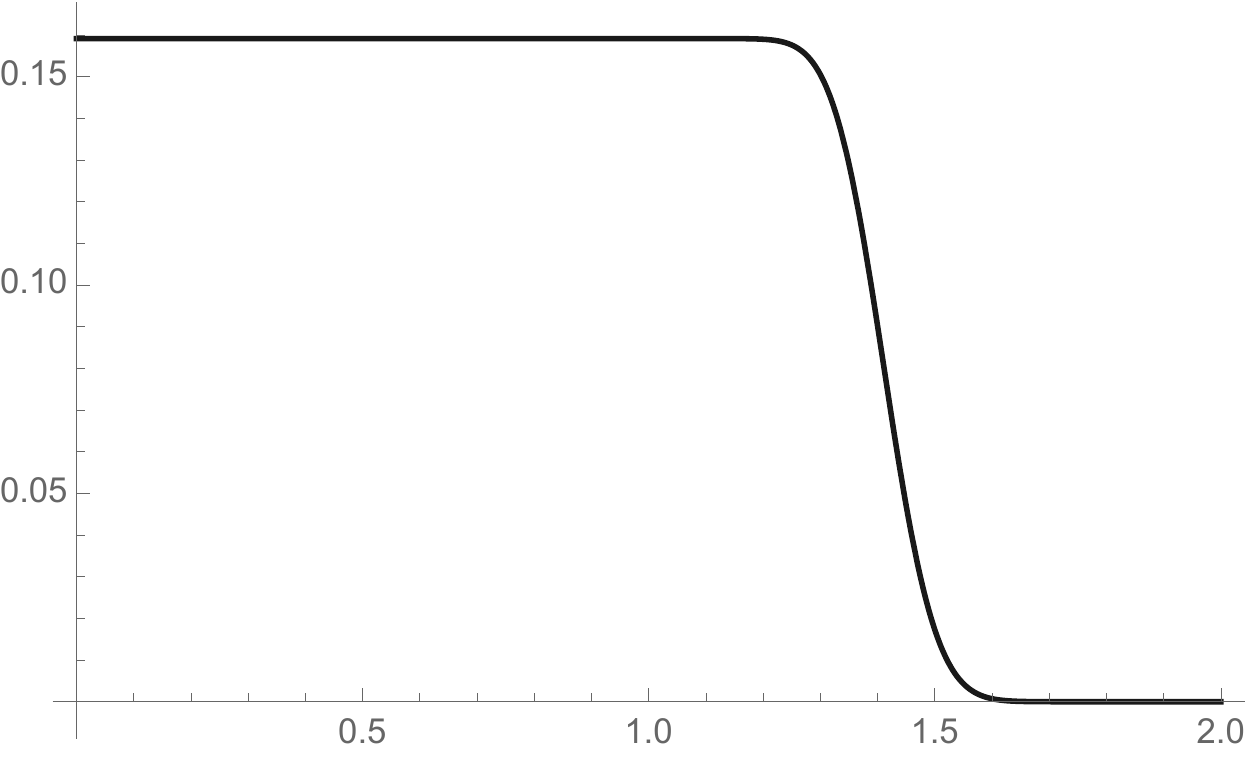}
   \hspace{4cm}
   \includegraphics[scale=0.4, bb=0 0 259 214]{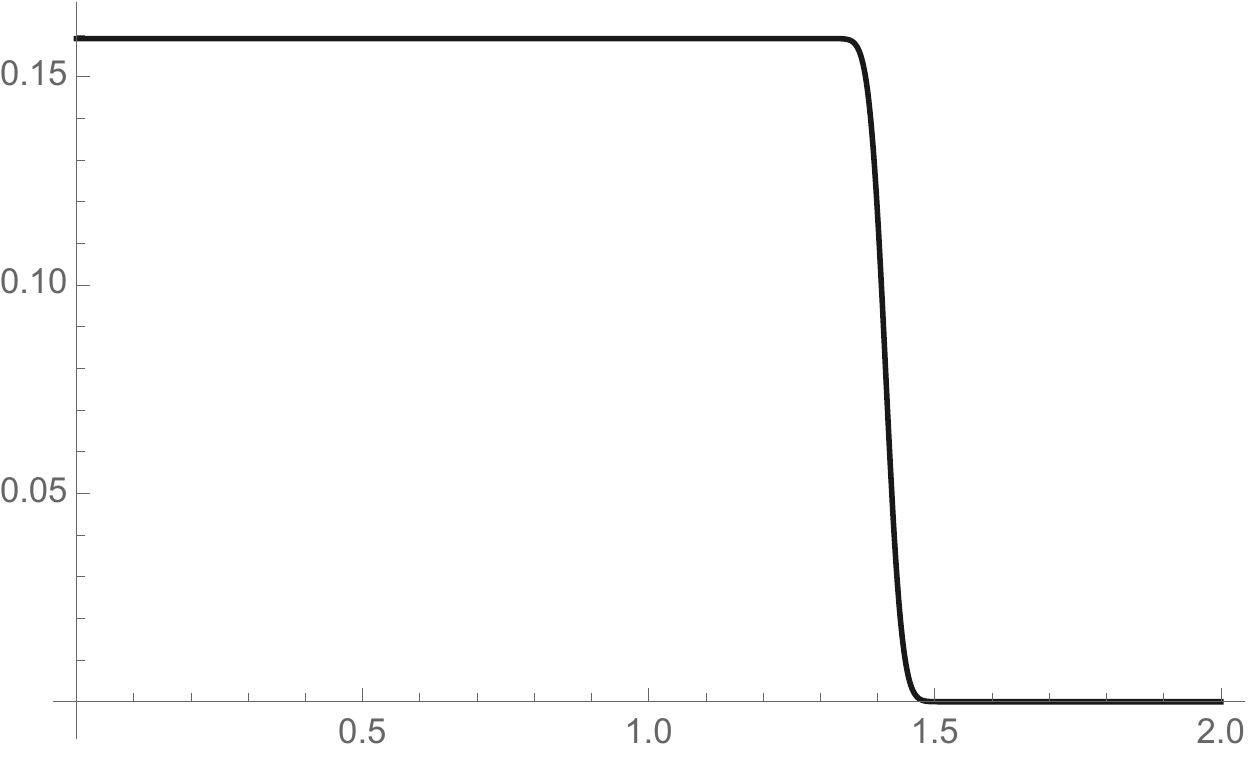}  
    \caption{Plot of the density function for $N=100$ (left)
    and $N=1000$ (right) with $N\theta =1$. 
    Their radii are $\simeq \sqrt{2N\theta}$. } 
    \label{density_sharpened}
  \end{center}
\end{figure}

Since the radius of the fuzzy disc is almost $\sqrt{2N\theta}$, 
the radii are about $0.44, 0.63, 0.77$ for $N=1, 2, 3$ and $\theta =0.1$.
The edge of the fuzzy disc becomes shaper as $N \to \infty$ with 
$N\theta =$ fixed.  In Fig.\ref{density_sharpened}, we draw $N=100$ and $N=1000$ 
with $N\theta = 1$ cases, respectively.
We can see that their radii are almost $\sqrt{2N\theta} \simeq 1.4$ in Fig.\ref{density_sharpened}. 

\begin{figure}[t]
  \begin{center}
   \includegraphics[scale=0.6, bb=0 0 300 215]{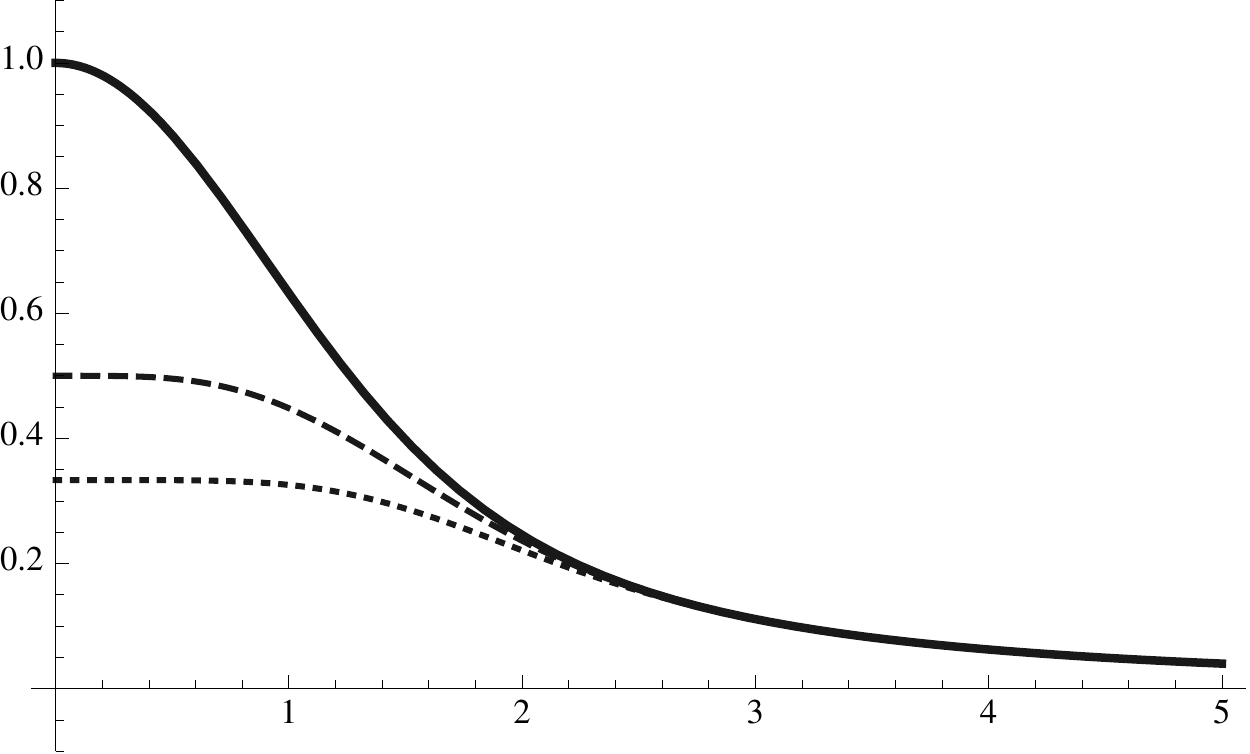}
    \caption{Plot of $h_{\sss N}^{\sss FD}(x)$ for 
    $N=1$ (solid), $N=2$ (dashed) and $N=3$ (dotted).} 
    \label{hFD}
  \end{center}
\end{figure}

Using the mass function
\ba
m_{\sss N}^{\sss FD}(r)
&=& 2\pi \int_0^r dr' r' \rho_N(r')\\ 
&=&\frac{M}{\Gamma(N+1)}\left[
\gamma\left(N+1,\frac{r^2}{2\theta}\right)
+\frac{r^2}{2\theta}\Gamma\left(N,\frac{r^2}{2\theta}\right)
\right], 
\ea
we explicitly write the horizon formation condition for the fuzzy disc as 
\be
m_{\sss N}^{\sss FD}(r_h) \geq \frac{r_h^2}{8M\ell^2}.
\ee
Introducing $x = r_h/\sqrt{2\theta}$, this condition is rewritten as
\be
h_{\sss N}^{\sss FD}(x) = 
\frac{1}{x^2}\left[1-\frac{\Gamma(N+1, x^2)}{\Gamma(N+1)}\right]
+\frac{\Gamma(N, x^2)}{\Gamma(N+1)}
\geq \frac{(\sqrt{2\theta})^2}{8M\ell^2}.
\ee
The profiles of $h_{\sss N}^{\sss FD}(x)$ for $N=1, 2, 3$ are 
shown in Fig.\ref{hFD}. 
Here $N$ denotes how many annuli are summed. 
The fuzzy disc with $N=1$ is constituted of $\ket{0}\bra{0}$ only, 
and the fuzzy disc with $N=2$ is the sum of two annuli that corresponds to 
$\ket{0}\bra{0}+\ket{1}\bra{1}$, ..., and so on. 

If we choose an appropriate $(\sqrt{2\theta})^2/(8M\ell^2)$,
there can exist a black hole. For any $N$, as the characteristic function 
$h_{\sss N}^{\sss FD}(x) $ is 
monotonically decreasing, we find 
\be
\begin{cases}
M \geq \diss \frac{(\sqrt{2\theta})^2}{8h_{\sss N}^{*}\ell^2} & : \mbox{one horizon}, 
\vspace{0.5cm}\\
M < \diss \frac{(\sqrt{2\theta})^2}{8h_{\sss N}^{*}\ell^2} & : \mbox{no horizon},
\end{cases}
\ee
where $h_{\sss N}^{*}$ is the maximum of $h_{\sss N}^{\sss FD}(x)$. 

The Ricci scalar of this spacetime is positive at $r=0$ for any $N$, 
which means that there is a de Sitter core there. 
This is same as the three-dimensional black hole 
with $\rho_0(r) \propto e^{-r^2/(2\theta)}$ as its source. 
Since the fuzzy disc source does not have a void in its center, 
its shape is similar to that described by $\rho_0(r)$. 
So this is reasonable.

\subsection{Extension to a four-dimensional black hole}

It is interesting to extend this fuzzy disc source 
to a four-dimensional spacetime. 
This extension corresponds 
to the source that is a sum of the thick matter layers considered in \cite{Nicolini:2011fy} 
with giving certain weights to the layers.  
For a three-dimensional case with the fuzzy disc as its source, 
two horizons can not be formed  
as we expected from the fact that there is no void around the centers. 
However, in four dimensions, the situation will change. 
In four dimensions, we consider the following density
motivated by the fuzzy ``disc'',
\be
\rho_{\sss N}^{\sss 4dFD}(r)
=\frac{3M}{4\pi (\sqrt{2\theta})^3 \Gamma(N+\frac{3}{2})}
\Gamma\left(N, \frac{r^2}{2\theta}\right),
\ee
and the mass function counterpart is given by
\be
m_{\sss N}^{\sss 4dFD}(r)
= \frac{M}{3\Gamma\left(N+\frac{3}{2}\right)}
\left[
\gamma\left(N+\frac{3}{2}, \frac{r^2}{2\theta}\right)
+\frac{r^3}{(\sqrt{2\theta})^3} \Gamma\left(N, \frac{r^2}{2\theta}\right)
\right].
\ee
The horizon formation condition is determined by 
\be
h_{\sss N}^{\sss 4dFD}(x)
=\frac{2}{3\Gamma(N+\frac{3}{2})}
\left[
\frac{1}{x}\gamma\left(N+\frac{3}{2}, x^2\right) + x^2 \Gamma(N, x^2)
\right],
\ee 
where $x=r_h/\sqrt{2\theta}$ as before. 
\begin{figure}[t]
  \begin{center}
   \includegraphics[scale=0.6, bb=0 0 300 214]{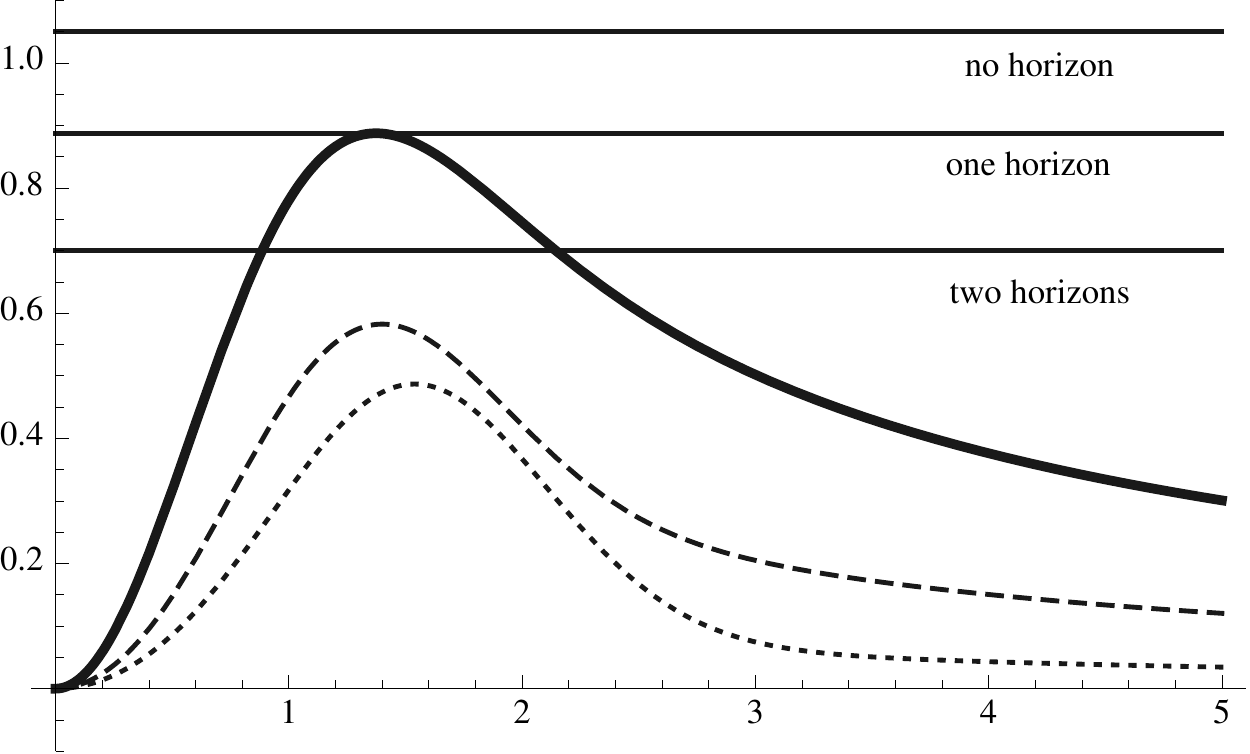}
    \caption{Plot of $h_{\sss N}^{\sss 4dFD}(x)$ 
    for $N=1$ (solid), $N=2$ (dashed) and $N=3$ (dotted). 
    The horizontal lines denote the values of $(\sqrt{2\theta})^2/8M\ell^2$.
    Though we draw the horizon formation condition only for $N=1$, 
    the qualitative behavior does not change for an arbitrary $N$. 
    In four dimensions, there is a black hole that can have two horizons
    if it holds an appropriate condition.  } 
    \label{h4dFD}
  \end{center}
\end{figure}
The behavior of $h_{\sss N}^{\sss 4dFD}(x)$ is shown in Fig.\ref{h4dFD}. 
As we expected, $h_{\sss N}^{\sss 4dFD}(x)$ has only one extremum at 
finite $x$. $h_{\sss N}^{\sss 4dFD}(x)$ asymptotically approaches to zero when $x \to \infty$
and  it becomes zero at $x=0$. 
We find that there is a black hole that can have two horizons as long as 
there is enough mass $M$ within a given volume. 
This is because the power of divergence of 
the characteristic function around $x=0$ is weakened from $x^{-2}$ in three dimensions
to $x^{-1}$ in four dimensions. 

We can conclude that there are three cases, that is, two horizons, one horizon (the extremal) 
and no horizon cases, respectively. For a black hole with two horizons, 
there would be a remnant after radiation from the black holes. 
It is true that the behaviors of the source terms depend on noncommutativity 
which is denoted by $\theta$, but we may have to say that 
the possibility of remnant is originated from the difference of dimensions 
and intrinsic structures of the spacetimes rather than noncommutativity.

\resection{Conclusion and discussion}

In this paper, we considered the black holes 
in three and four-dimensional spacetimes. 
These black holes have the fuzzy sources inspired by noncommutative geometry. 
Noncommutativity between space coordinates 
is translated to the Gaussian profiles of matter distributions 
represented by the noncommutative parameter $\theta$.

As investigated by many authors, 
there can be a black hole with two horizons  for such a source
when enough mass is included within a given radius. 
In order to judge whether a horizon is formed or not, 
we introduced the mass functions. 
Introduction of them makes it possible 
to regard those black holes as Schwarzschild black holes 
with effective masses.

As an example, we first showed that 
how the mass function effectively works to investigate 
the horizon formation condition in the four-dimensional case 
with the density distribution represented by the generalized Gaussian function 
argued in \cite{Nicolini:2005vd, Nicolini:2011fy}. 
Next we applied this manner to the three-dimensional spacetime
with the source described by the generalized Gaussian function. 
In the case of the three-dimensional spacetimes,  
the horizon formation condition depends on 
whether the mass function evaluated at a given radius 
is larger than the mass of the BTZ black hole. 
We analyzed the behaviors of the characteristic function
for the horizon formation condition in detail
and found how the difference between the three- and the four-dimensional spacetimes
affects the horizon formation condition. 
The essential point of the horizon formation is 
the existence of a void around the center of the spacetime, 
which is closely relates to the spacetime's dimension
rather than noncommutativity that is expected to work as a repulsive force
as a quantum effect. 
We saw this by giving the toy model that is apart from noncommutative 
geometry inspired models. 
 
Since our point of view by means of mass function and characteristic 
function is graphical and intuitively understandable, 
we can easily apply it to any sources. 
In fact, we also considered the black hole 
with the source whose density distribution is motivated by the fuzzy disc. 
For such a fuzzy disc source with an arbitrary radius, 
a black hole can be formed as long as enough mass is included inside a given circle. 
This behavior is similar to the three-dimensional black hole with 
the density distribution $\rho \propto e^{-r^2/(2\theta)}$. 
This is interpreted as both distributions do not have a void
at the centers of the spacetimes, but have de Sitter cores. 
The only difference between them is the length of the plateaus from the origin.  
We also considered the sources that have the same profile as the fuzzy disc 
in four dimensions. Since the fuzzy ``disc" is an three-dimensional object, 
it is just a toy model to check how the difference of dimensions works 
on the horizon formation condition. 
In four dimensions, there can exist a black hole with two horizons 
for the source whose density distribution is motivated by the fuzzy disc.

As for a void, we want to state that it might be interesting 
to consider a source whose density distribution has the same 
profile as the fuzzy annulus we found in \cite{Kobayashi:2012ek, Kobayashi:2015hea}. 
The density distribution of the fuzzy annulus can be written by 
the linear combination of arbitrary number of the generalized Gaussian functions. 
According the argument so far, we expect that there could 
exist a black hole with two horizons even in three dimensions. 
Furthermore, there could be a black hole with more than two horizons 
because it is possible to put any gap between annuli. 
It is worth while analyzing the interior structure of such a spacetime 
relating to the Hawking radiation as a probe \cite{Deng:2016qua}, 
which is left for a sequent paper. 
Also, the detail analysis on causal structure, geodesic motion of a particle, 
thermodynamics and so on would  also be interesting.

\section*{Acknowledgments}
We would like to thank to T. Asakawa and D. Ida for fruitful discussions.


\bibliographystyle{JHEP}
\bibliography{refsA}

\end{document}

%% file: promise.tex
\newcommand{\be}{\begin{equation}}
\newcommand{\ee}{\end{equation}}
\newcommand{\ba}{\begin{eqnarray}}
\newcommand{\ea}{\end{eqnarray}}
\newcommand{\nn}{\nonumber}
\newcommand{\qq}{\qquad}
\newcommand{\del}{\partial}
\newcommand{\bra}[1]{\left\langle\,{#1}\,\right|}
\newcommand{\ket}[1]{\left|\,{#1}\,\right\rangle}
\newcommand{\bracket}[2]{
\left\langle\left.\,{#1}\,\right|\,{#2}\,\right\rangle}
\newcommand{\eq}[1]{(\ref{#1})}

\newcommand{\vecvar}[1]{\mbox{\boldmath$#1$}}
\newcommand{\diss}{\displaystyle}

\newcommand{\bz}{\overline{z}}
\newcommand{\tb}{\widetilde{b}}
\newcommand{\tc}{\widetilde{c}}
\newcommand{\talpha}{\widetilde{\alpha}}
\newcommand{\tD}{\widetilde{D}}
\newcommand{\tB}{\widetilde{B}}
\newcommand{\tF}{\widetilde{F}}
\newcommand{\tG}{\widetilde{G}}
\newcommand{\bb}{\overline{b}}
\newcommand{\bc}{\overline{c}}
\newcommand{\br}{\overline{r}}
\newcommand{\bq}{\overline{q}}
\newcommand{\bep}{\overline{\epsilon}}
\newcommand{\bN}{\bar{N}}
\newcommand{\calO}{{\mathcal{O}}}
\newcommand{\hO}{\hat{O}}
\newcommand{\hh}{\widehat{h}}
\newcommand{\hD}{\widehat{D}}
\newcommand{\tL}{\widetilde{L}}
\newcommand{\mbb}{{\mathbf b}}
\newcommand{\dpsl}{\displaystyle}
\newcommand{\prpr}{\prime\prime}
\newcommand{\tPsi}{\widetilde{\Psi}}
\newcommand{\bX}{{\mathbf X}}
\newcommand{\bD}{\overline{D}}
\newcommand{\bTheta}{\overline{\Theta}}
\newcommand{\bw}{\bar{w}}
\newcommand{\bP}{\mathbf P}
\newcommand{\bM}{\mathbf M}
\newcommand{\cD}{{\cal{D}}}
\newcommand{\bGamma}{\mathbf \Gamma}

\newcommand{\hx}{\hat{x}}
\newcommand{\hpsi}{\hat{\psi}}
\newcommand{\hphi}{\hat{\phi}}
\newcommand{\hp}{\hat{p}}
\newcommand{\hpi}{\hat{\pi}}
\newcommand{\ta}{\tilde{a}}
\newcommand{\hX}{\hat{X}}
\newcommand{\hP}{\hat{P}}

\newcommand{\hPsi}{\hat{\Psi}}
\newcommand{\hPhi}{\hat{\Phi}}
\newcommand{\hPi}{\hat{\Pi}}
\newcommand{\hsigma}{\hat{\sigma}}
\newcommand{\hT}{\hat{T}}
\newcommand{\bx}{\mathbf x}
\newcommand{\bp}{\mathbf p}
\newcommand{\hN}{\hat{N}}

\newcommand{\tr}{\rm tr}
\newcommand{\Tr}{{\rm Tr}}
\newcommand{\Z}{{\mathbb Z}}

\newcommand{\vtr}{\vartriangleright}

\newcommand{\ep}{\epsilon}
\newcommand{\Est}{E^{\star}}
\newcommand{\Vst}{V_{\star}}
\newcommand{\st}{\star}

\newcommand{\hU}{\hat{U} }
\newcommand{\PB}{\hat{U} _{PB}}
\newcommand{\daga}{\hat{a}^{\dagger}}
\newcommand{\ha}{\hat{a}}

\newcommand{\sst}{\scriptstyle}
\newcommand{\sss}{\scriptscriptstyle}

%% file: paper_A07_arXivRevised.bbl
\providecommand{\href}[2]{#2}\begingroup\raggedright\begin{thebibliography}{10}

\bibitem{Aschieri:2005zs}
P.~Aschieri, M.~Dimitrijevic, F.~Meyer and J.~Wess, {\it {Noncommutative
  geometry and gravity}},  {\em Class. Quant. Grav.} {\bf 23} (2006) 1883--1912
  [\href{http://arXiv.org/abs/hep-th/0510059}{{\tt hep-th/0510059}}].

\bibitem{Aschieri:2005yw}
P.~Aschieri {\em et.~al.}, {\it {A gravity theory on noncommutative spaces}},
  {\em Class. Quant. Grav.} {\bf 22} (2005) 3511--3532
  [\href{http://arXiv.org/abs/hep-th/0504183}{{\tt hep-th/0504183}}].

\bibitem{Aschieri:2009qh}
P.~Aschieri and L.~Castellani, {\it {Noncommutative Gravity Solutions}},  {\em
  J. Geom. Phys.} {\bf 60} (2010) 375--393
  [\href{http://arXiv.org/abs/0906.2774}{{\tt 0906.2774}}].

\bibitem{Asakawa:2009yb}
T.~Asakawa and S.~Kobayashi, {\it {Noncommutative Solitons of Gravity}},  {\em
  Class. Quant. Grav.} {\bf 27} (2010) 105014
  [\href{http://arXiv.org/abs/0911.2136}{{\tt 0911.2136}}].

\bibitem{Kobayashi:2009baa}
S.~Kobayashi and T.~Asakawa, {\it {Emergence of Spacetimes and
  Noncommutativity}},  in {\em {Proceedings, 19th Workshop on General
  Relativity and Gravitation in Japan (JGRG19), Tokyo, Japan}}, 2009.

\bibitem{Nicolini:2005vd}
P.~Nicolini, A.~Smailagic and E.~Spallucci, {\it {Noncommutative geometry
  inspired Schwarzschild black hole}},  {\em Phys. Lett.} {\bf B632} (2006)
  547--551 [\href{http://arXiv.org/abs/gr-qc/0510112}{{\tt gr-qc/0510112}}].

\bibitem{Ansoldi:2006vg}
S.~Ansoldi, P.~Nicolini, A.~Smailagic and E.~Spallucci, {\it {Noncommutative
  geometry inspired charged black holes}},  {\em Phys. Lett.} {\bf B645} (2007)
  261--266 [\href{http://arXiv.org/abs/gr-qc/0612035}{{\tt gr-qc/0612035}}].

\bibitem{Nicolini:2009gw}
P.~Nicolini and E.~Spallucci, {\it {Noncommutative geometry inspired wormholes
  and dirty black holes}},  {\em Class. Quant. Grav.} {\bf 27} (2010) 015010
  [\href{http://arXiv.org/abs/0902.4654}{{\tt 0902.4654}}].

\bibitem{Nicolini:2011fy}
P.~Nicolini, A.~Orlandi and E.~Spallucci, {\it {The final stage of
  gravitationally collapsed thick matter layers}},  {\em Adv. High Energy
  Phys.} {\bf 2013} (2013) 812084 [\href{http://arXiv.org/abs/1110.5332}{{\tt
  1110.5332}}].

\bibitem{Spallucci:2008ez}
E.~Spallucci, A.~Smailagic and P.~Nicolini, {\it {Pair creation by higher
  dimensional, regular, charged, micro black holes}},  {\em Phys. Lett.} {\bf
  B670} (2009) 449--454 [\href{http://arXiv.org/abs/0801.3519}{{\tt
  0801.3519}}].

\bibitem{Smailagic:2010nv}
A.~Smailagic and E.~Spallucci, {\it {'Kerrr' black hole: the Lord of the
  String}},  {\em Phys. Lett.} {\bf B688} (2010) 82--87
  [\href{http://arXiv.org/abs/1003.3918}{{\tt 1003.3918}}].

\bibitem{Modesto:2010rv}
L.~Modesto and P.~Nicolini, {\it {Charged rotating noncommutative black
  holes}},  {\em Phys. Rev.} {\bf D82} (2010) 104035
  [\href{http://arXiv.org/abs/1005.5605}{{\tt 1005.5605}}].

\bibitem{Nicolini:2008aj}
P.~Nicolini, {\it {Noncommutative Black Holes, The Final Appeal To Quantum
  Gravity: A Review}},  {\em Int. J. Mod. Phys.} {\bf A24} (2009) 1229--1308
  [\href{http://arXiv.org/abs/0807.1939}{{\tt 0807.1939}}].

\bibitem{Mureika:2011py}
J.~R. Mureika and P.~Nicolini, {\it {Aspects of noncommutative
  (1+1)-dimensional black holes}},  {\em Phys. Rev.} {\bf D84} (2011) 044020
  [\href{http://arXiv.org/abs/1104.4120}{{\tt 1104.4120}}].

\bibitem{Larranaga:2014uca}
A.~Larranaga, A.~Cardenas-Avendano and D.~A. Torres, {\it {On a general class
  of regular rotating black holes based on a smeared mass distribution}},  {\em
  Phys. Lett.} {\bf B743} (2015) 492--502
  [\href{http://arXiv.org/abs/1410.0049}{{\tt 1410.0049}}]. [Erratum: Phys.
  Lett.B747,564(2015)].

\bibitem{Dymnikova:1992ux}
I.~Dymnikova, {\it {Vacuum nonsingular black hole}},  {\em Gen. Rel. Grav.}
  {\bf 24} (1992) 235--242.

\bibitem{Dymnikova:2003vt}
I.~Dymnikova, {\it {Spherically symmetric space-time with the regular de Sitter
  center}},  {\em Int. J. Mod. Phys.} {\bf D12} (2003) 1015--1034
  [\href{http://arXiv.org/abs/gr-qc/0304110}{{\tt gr-qc/0304110}}].

\bibitem{Dymnikova:2004qg}
I.~Dymnikova and E.~Galaktionov, {\it {Stability of a vacuum nonsingular black
  hole}},  {\em Class. Quant. Grav.} {\bf 22} (2005) 2331--2358
  [\href{http://arXiv.org/abs/gr-qc/0409049}{{\tt gr-qc/0409049}}].

\bibitem{Rahaman:2013gw}
F.~Rahaman, P.~K.~F. Kuhfittig, B.~C. Bhui, M.~Rahaman, S.~Ray and U.~F.
  Mondal, {\it {BTZ black holes inspired by noncommutative geometry}},  {\em
  Phys. Rev.} {\bf D87} (2013), no.~8 084014
  [\href{http://arXiv.org/abs/1301.4217}{{\tt 1301.4217}}].

\bibitem{Tejeiro:2010gu}
J.~M. Tejeiro and A.~Larranaga, {\it {Noncommutative Geometry Inspired Rotating
  Black Hole in Three Dimensions}},  {\em Pramana} {\bf 78} (2012) 155--164
  [\href{http://arXiv.org/abs/1004.1120}{{\tt 1004.1120}}].

\bibitem{Larranaga:2010tt}
A.~Larranaga and J.~M. Tejeiro, {\it {Three Dimensional Charged Black Hole
  Inspired by Noncommutative Geometry}},  {\em Abraham Zelmanov J.} {\bf 4}
  (2011) 28--35 [\href{http://arXiv.org/abs/1004.1608}{{\tt 1004.1608}}].

\bibitem{Rahaman:2014pha}
F.~Rahaman, P.~Bhar, R.~Sharma and R.~K. Tiwari, {\it {Noncommutative geometry
  inspired $3$-dimensional charged black hole solution in an anti-de Sitter
  background spacetime}},  {\em Eur. Phys. J.} {\bf C75} (2015), no.~3 107
  [\href{http://arXiv.org/abs/1409.0552}{{\tt 1409.0552}}].

\bibitem{Myung:2008kp}
Y.~S. Myung and M.~Yoon, {\it {Regular black hole in three dimensions}},  {\em
  Eur. Phys. J.} {\bf C62} (2009) 405--411
  [\href{http://arXiv.org/abs/0810.0078}{{\tt 0810.0078}}].

\bibitem{Jun:2014jqa}
J.~Liang, Y.-C. Liu and Q.~Zhu, {\it {Thermodynamics of noncommutative geometry
  inspired black holes based on Maxwell-Boltzmann smeared mass distribution}},
  {\em Chin. Phys.} {\bf C38} (2014) 025101.

\bibitem{Poisson}
E.~Poisson, {\em A Relativist's Toolkit: The Mathematics of Black-Hole
  Mechanics}.
\newblock Cambridge University Press, 2007.

\bibitem{Spallucci:2009zz}
E.~Spallucci, A.~Smailagic and P.~Nicolini, {\it {Non-commutative geometry
  inspired higher-dimensional charged black holes}},  {\em Phys. Lett.} {\bf
  B670} (2009) 449--454 [\href{http://arXiv.org/abs/0801.3519}{{\tt
  0801.3519}}].

\bibitem{Park:2008ud}
M.-I. Park, {\it {Smeared hair and black holes in three-dimensional de Sitter
  spacetime}},  {\em Phys. Rev.} {\bf D80} (2009) 084026
  [\href{http://arXiv.org/abs/0811.2685}{{\tt 0811.2685}}].

\bibitem{Yamazaki:2001ue}
R.~Yamazaki and D.~Ida, {\it {Black holes in three-dimensional
  Einstein-Born-Infeld dilaton theory}},  {\em Phys. Rev.} {\bf D64} (2001)
  024009 [\href{http://arXiv.org/abs/gr-qc/0105092}{{\tt gr-qc/0105092}}].

\bibitem{Kobayashi:2012ek}
S.~Kobayashi and T.~Asakawa, {\it {Angles in Fuzzy Disc and Angular
  Noncommutative Solitons}},  {\em JHEP} {\bf 04} (2013) 145
  [\href{http://arXiv.org/abs/1206.6602}{{\tt 1206.6602}}].

\bibitem{Lizzi:2003ru}
F.~Lizzi, P.~Vitale and A.~Zampini, {\it {The fuzzy disc}},  {\em JHEP} {\bf
  08} (2003) 057 [\href{http://arXiv.org/abs/hep-th/0306247}{{\tt
  hep-th/0306247}}].

\bibitem{Lizzi:2003hz}
F.~Lizzi, P.~Vitale and A.~Zampini, {\it {From the fuzzy disc to edge currents
  in Chern-Simons theory}},  {\em Mod. Phys. Lett.} {\bf A18} (2003) 2381--2388
  [\href{http://arXiv.org/abs/hep-th/0309128}{{\tt hep-th/0309128}}].

\bibitem{Lizzi:2006bu}
F.~Lizzi, P.~Vitale and A.~Zampini, {\it {The fuzzy disc: A review}},  {\em J.
  Phys. Conf. Ser.} {\bf 53} (2006) 830--842.

\bibitem{Kobayashi:2015hea}
S.~Kobayashi and T.~Asakawa, {\it {Fuzzy Objects and Noncommutative Solitons}},
   in {\em {Proceedings, 13th Marcel Grossmann Meeting on Recent Developments
  in Theoretical and Experimental General Relativity, Astrophysics, and
  Relativistic Field Theories (MG13), Stockholm, Sweden}}, pp.~2522--2524,
  2015.

\bibitem{Deng:2016qua}
Y.~Deng and G.~Cleaver, {\it {Hawking Radiation from Regular Black Hole as a
  Possible Probe for Black Hole Interior Structure}},
  \href{http://arXiv.org/abs/1602.06035}{{\tt 1602.06035}}.

\end{thebibliography}\endgroup
